\RequirePackage{ifpdf}
\documentclass[a4paper]{JHEP3}
\usepackage{epsfig}
\usepackage{amsmath}
\usepackage{graphicx}
\usepackage{amssymb}
\usepackage{latexsym}

\newcommand{\be}{\begin{equation}}
\newcommand{\ee}{\end{equation}}
\newcommand{\bea}{\begin{eqnarray}}
\newcommand{\eea}{\end{eqnarray}}

\title{Killing vectors of FLRW\footnote{We thank Luc Blanchet for insisting that we use FLRW instead of FRW.} metric (in comoving coordinates) and zero modes of the scalar Laplacian .}
\author{N.D. Hari Dass
\\  Visiting Professor, TIFR-TCIS, Hyderabad 500075\\
 Email: \email{dass@tifrh.res.in}}
\author{Harini Desiraju
\\Erasmus Exchange Student, Physics Department, Uppsala University, and,
\\Physics Department, University of Hyderabad\\
Email:\email{harini.desiraju@gmail.com}}

\abstract{Based on an examination of the solutions to the Killing \cite{killingorig} Vector equations for the 
FLRW-metric in co moving coordinates
\cite{friedmanorig, lemaitre, robertsonorig, walkerorig}, it is conjectured and proved
that the components(in these coordinates) of Killing Vectors, when suitably scaled by functions, are \emph{zero modes}
of the corresponding
\emph{scalar} Laplacian. The complete such set of zero modes(infinitely many) are explicitly constructed
for the two-sphere. They are parametrised by an integer n. For $n\,\ge\,2$, all the solutions are \emph{irregular}
(in the sense that they are neither well defined everywhere nor are \emph{square-integrable}). The associated 2-d
vectors are also \emph{not normalisable}. The $n=0$ solutions being constants (these correspond to the zero angular momentum
solutions) are regular and normalizable. Not all of the $n=1$ solutions are regular but the associated vectors are normalizable.
Of course, the action of scalar Laplacian coordinate independent significance only when acting on scalars. 
However, our conclusions have an unambiguous meaning as long as one works in this coordinate system. 
As an intermediate step, the covariant Laplacians(vector Laplacians) of Killing vectors are worked out for four-manifolds in two different
ways, both of which have the novelty of not explicitly needing the connections. It is further shown that for certain maximally symmetric 
sub-manifolds(hypersurfaces of one or more constant comoving coordinates) of the FLRW-spaces also, the scaled
Killing vector components are zero modes of their corresponding scalar Laplacians. The Killing vectors for the maximally symmetric 
four-manifolds are worked out using the elegant embedding 
formalism originally due to Schr\"odinger \cite{schrobook, weinbook}. 
Some consequences of our results are worked out. Relevance to some very recent works on zero modes
in AdS/CFT correspondences \cite{keelerng,belinmaloney}, as well as on braneworld scenarios 
\cite{durrerkocian,giovannini} is briefly commented upon.
}


\begin{document}

\section{Introduction}
Manifolds are the arena where the dynamical laws of physics play themselves out. They are also objects of the most extensive and fundamental
of mathematical studies. Themes of central importance to both the physicist and the mathematician, are the symmetry aspects of such manifolds.
While for the physicist symmetry has such diverse ramifications such as conservation laws, solutions of the dynamical evolution equations etc, 
for the mathematician the chief challenge is extracting the geometrical content of the symmetries of manifolds. In particular, to be able to
make statements about the symmetries of manifolds that are independent of any choice of coordinates that may have been made. The point is that while symmetries may sometimes be manifest in certain choices of coordinates, they may be completely obscured in certain other choices of coordinates. An illuminating example is that of flat Euclidean space, say, in three dimensions. The symmetries in question are the three translations
and three rotations. While in the cartesian system of coordinates, all these are manifest, in spherical polar coordinates the translations are
not obvious as symmetries. A particularly useful concept is that of the \emph{maximally symmetric spaces}. It can be proved that a D-dimensional manifold can at most have $\frac{D(D+1)}{2}$ isometries(Killing vectors), and the space with this maximal number of isometries is called a
maximally symmetric space. For a good exposition of these concepts see \cite{weinbook,hartlebook}. More advanced treatments can be found in
\cite{eisenhart1, eisenhart2,helgason}. De Sitter spaces \cite{desitter} which have played important roles in cosmology \cite{weinbook,hartlebook,liddlebook,carroll} for a long time, and which have again become its central themes, are examples of such spaces. More realistic cosmological
models are of the FLRW-type \cite{friedmanorig,lemaitre,robertsonorig,walkerorig}.

A systematic coordinate-independent  way of addressing such issues is through the so called \emph{isometries} of the manifold. The 
infinitesimal isometries are characterised by the so-called Killing vectors and they are governed by a set of partial differential equations 
called Killing vector equations. These equations are remarkably restrictive and they completely determine the Killing vectors, at least
locally. In what follows, we explicitly write down the Killing vector equations for the FLRW-metric. Solving the Killing vector equations can be
a tedious exercise, needing some degree of ingenuity. We shall simply write down the solutions for the case of the FLRW  cosmology. This is
a case where the four-manifold is not maximally symmetric. In the (t,r,$\theta,\phi$) parametrisation, the $t = const.$ sections are, however,
maximally symmetric. Likewise, every subspace of this 3-space with one or more comoving coordinates is maximally symmetric.

We observed a curiosity as far as the Killing vectors of the $\theta,\phi$ subspace i.e the 2-sphere were concerned: denoting the Killing
vector components for this case by $\xi^\theta, \xi^\phi$, it was found that both $\xi^\phi, \frac{\xi^\theta}{\sin{\theta}}$ were eigenmodes of the
scalar Laplacian on the 2-sphere with zero eigenvalues. Furthermore, some of them were ill defined at $\theta = 0, \pi$. At the same time, they
were also not square integrable. These two features are logically independent as there are functions which are not well defined everywhere and
yet are square-integrable. Extreme care is needed While discussing whether functions are well defined everywhere or not, as the $\theta,\phi$ coordinate
system breaks down precisely at $\theta\,=\,0,\pi$. That the Laplacian involved was the
scalar Laplacian instead of the covariant, or in this case the vector Laplacian \cite{weinbook}, came as a surprise. In fact, it will be shown during the 
course of this article that covariant Laplacians on Killing vectors do not vanish except in Ricci-flat spaces. The fact that $\xi^\theta$ had to
be divided by $\sin{\theta}$ while $\xi^\phi$ required no such scaling further intrigued us.

On the basis of this observation we made the conjecture that  all components of the most general Killing vector of maximally 
symmetric spaces, after scaling by a function that depends on which component we take, are zero modes of the scalar Laplacians 
on these manifolds. In the subsequent sections, we prove this conjecture for the 
$t=constant$ hypersurfaces of the space with FLRW metric and for the maximally symmetric four-manifolds also with FLRW metric.
\section{Isometries and Killing vectors}
Under a general coordinate transformation
\be
\label{eq:gct}
x^\mu\,\rightarrow\,{x^\prime}^\mu
\ee
the metric $g_{\mu\nu}$ transforms as
\be
\label{eq:metrictrans}
g^\prime(x^\prime)_{\mu^\prime\nu^\prime} = 
\frac{\partial\,x^\mu}{\partial {x^\prime}^{\mu^\prime}}
\frac{\partial\,x^\nu}{\partial {x^\prime}^{\nu^\prime}}\:g_{\mu\nu}(x)
\ee
Such coordinate transformations generically change some or all of the metric components. The so called \emph{isometries}, however,
do not change the form of the metric:
\be
\label{eq:isometry}
g^\prime(x^\prime)_{\mu\nu} = g(x)_{\mu\nu}
\ee
Of special interest are the so called infinitesimal isometries
\be
\label{eq:kvectors}
{x^\prime}^\mu = x^\mu +\epsilon\,\xi^\mu
\ee
where $\epsilon$ denotes a small number. The vectors $\xi^\mu$ are called the Killing vectors.
\subsection{Killing vector equations}
It is a straightforward consequence of eqn.(\ref{eq:isometry}) that the Killing vectors obey
\be
\label{eq:killingeqns}
\xi_{\mu;\nu}+\xi_{\nu;\mu} =0
\ee
where $V_{\mu;\nu}$ stand for the covariant derivatives
\be
\label{eq:covderivs}
V_{\mu;\nu} = \frac{\partial\,V_\mu}{\partial\,x^\nu}-\Gamma^\lambda_{\mu\nu}\,V_\lambda
\ee
For future use we also record the covariant derivatives of contravarant vectors $V^\lambda$:
\be
\label{eq:covderivcontra}
V^\mu_{\,;\nu} = \frac{\partial\,V^\mu}{\partial\,x^\nu}+\Gamma^\lambda_{\mu\nu}\,V_\lambda
\ee
In eqns.(\ref{eq:covderivs},\ref{eq:covderivcontra}), we have followed the conventions of Weinberg \cite{weinbook}, with
the Christoffel connection given by(see eqn 4.5.2 of \cite{weinbook}):
\be
\label{eq:christoffel}
\Gamma^\lambda_{\mu\nu} = \frac{1}{2}\,g^{\lambda\sigma}\big\{g_{\sigma\mu,\nu}+g_{\sigma\nu,\mu}-g_{\mu\nu.\sigma}\big\}
\ee
\section{FLRW metric: generalities}
The FLRW metric with our signature convention and in the so-called comoving coordinates takes the form
\be
\label{eq:frwmetric0}
ds^2 = dt^2 - R(t)^2\{\frac{dr^2}{1-kr^2}+r^2d\theta^2+r^2\sin^2\theta\,d\phi^2\}
\ee
Thus the metric $g_{\mu\nu}$ and the inverse metric $g^{\mu\nu}$ are respectively given by:
\be
\label{eq:metriccomps}
g_{tt} = 1\quad\quad g_{rr} = -\frac{R^2(t)}{1-kr^2}\quad\quad g_{\theta\theta} = -R^2(t)\,r^2\quad\quad 
g_{\phi\phi} = -R^2(t)\,r^2\,\sin^2\,\theta
\ee
\be
\label{eq:inversemetric}
g^{tt} =1\quad\quad g^{rr} = -\frac{1-kr^2}{R^2(t)}\quad\quad g^{\theta\theta} = - \frac{1}{R^2(t)\,r^2}\quad\quad
g^{\phi\phi} = -\frac{1}{R^2(t)\,r^2\,\sin^2\theta}
\ee
The negative determinant of the metric, ${\mathrm g}$ is given by
\be
\label{eq:detg}
-g = R^6(t)\,\frac{r^4}{1-kr^2}\,\sin^2\,\theta\,\ge\,0
\ee
\subsection{The Christoffel Connections for the FLRW metric}
\bea
\label{eq:frwconnections}
& &\Gamma^0_{11}= \frac{R{\dot R}}{1-kr^2}\quad \Gamma^0_{22} = R{\dot R}\,r^2\quad \Gamma^0_{33} = R{\dot R}r^2\sin^2\theta\nonumber\\
& &\Gamma^1_{01}=\Gamma^1_{10}=\Gamma^2_{02}=\Gamma^2_{20}=\Gamma^3_{03}=\Gamma^3_{30}=\frac{{\dot R}}{R}\nonumber\\
& &\Gamma^1_{11} = \frac{kr}{1-kr^2}\quad \Gamma^1_{22} = -r(1-kr^2)\quad \Gamma^1_{33} = -r(1-kr^2)\sin^2\theta\nonumber\\
& &\Gamma^2_{12}=\Gamma^2_{21}=\Gamma^3_{13}=\Gamma^3_{31}=\frac{1}{r}\nonumber\\
& &\Gamma^2_{33}=-\sin\theta\,\cos\theta \quad \Gamma^3_{23}=\Gamma^3_{32}=\cot\theta
\eea 
A good discussion can also be found in \cite{carroll}.
\subsection{The curvature tensor}
The definition of Riemann curvature tensor we are using (see Weinberg 6.6.2 \cite{weinbook}) is:\\
\be
\label{eq:curvaturetensor}
R_{\lambda\mu\nu\kappa} = \frac{1}{2}\,\big[\frac{\partial^2\,g_{\lambda\nu}}{\partial\,x^\kappa\,\partial\,x^\mu}-
\frac{\partial^2\,g_{\lambda\kappa}}{\partial\,x^\mu\,\partial\,x^\nu}+\dots\big ]+g_{\eta\sigma}\,\big [\Gamma^{\eta}_{\nu\lambda}\,\Gamma^{\sigma}_{\mu\kappa}+\ldots\big ]
\ee
The Ricci tensor is given by:
\be
\label{eq:ricci}
R_{\mu\kappa} = g^{\lambda\nu}\,R_{\lambda\mu\nu\kappa}
\ee
Finally, the \emph{mixed component} Ricci tensor is given by
\be
\label{eq:mixedricci}
R^\mu_{\,\kappa} = g^{\mu\nu}\,R_{\nu\kappa}
\ee
For the FLRW metric of eqn.(\ref{eq:frwmetric0}), the components of $R^\mu_\nu$ are easily calculated to be:
\bea
\label{eq:mixedriccivalues}
& & R^0_{\,\,0} = 3\,\frac{{\ddot R}}{R}\nonumber\\
& & R^i_{\,\,j} = \frac{R{\ddot R}+2\,{\dot R}^2+\,2k}{R^2}\,\delta_{ij}
\eea
A necessary and sufficient condition for the entire four-manifold to be maximally symmetric is given by
\be
\label{eq:maximal}
3\,\frac{{\ddot R}}{R} = \frac{R{\ddot R}+2\,{\dot R}^2+\,2k}{R^2}
\ee 
Hence, for generic cases of $R(t)$, only the constant time hypersurfaces are maximally symmetric. 
\subsection{The scalar Laplacian}
Because of its central role in this paper, the scalar Laplacian $L^{(s)}$ for the FLRW metric will be explicitly
derived here. By definition
\be
\label{eq:scalarlapdef}
L^{(s)}\:f = \frac{1}{\sqrt{g}}\:\partial_\mu\,\sqrt{g}\,g^{\mu\nu}\,\partial_\nu\,f
\ee
It is straightforward to show that
\bea
\label{eq:scalarlap}
L^{(s)}\,f &=& \partial_t^2\,f-\frac{1-kr^2}{R^2(t)}\,\partial_r^2\,f-\frac{1}{R^2(t)\,r^2}\,\partial_\theta^2\,f
-\frac{1}{R^2(t)\,r^2\,\sin^2\theta}\,\partial_\phi^2\,f\nonumber\\
& & +3\,\frac{{\dot R(t)}}{R(t)}\,\partial_t\,f
-\frac{1-kr^2}{R^2(t)}\,\left (\frac{2}{r}-\frac{kr}{1-kr^2}\right )\,\partial_r\,f
-\frac{\cot\,\theta}{R^2(t)\,r^2}\,\partial_\theta\,f
\eea 
For later use, it will prove efficient to split $L^{(s)}$ as
\be
\label{eq:scalarlapsplit}
L^{(s)} = L^{(s)}_t\,+L^{(s)}_r\,+L^{(s)}_\theta\,+L^{(s)}_\phi
\ee
where
\begin{subequations}
\label{eq:splitscalarlap}
\bea
L^{(s)}_t\,&=& \partial_t^2\,+3\,\frac{{\dot R}}{R}\,\partial_t\label{eq:scalarlap-t}\\
L^{(s)}_r\,&=& -\frac{1-kr^2}{R^2}\,\partial_r^2 -\frac{1}{R^2}\left (\frac{2(1-kr^2)}{r}\,-kr\right )\,\partial_r
\label{eq:scalarlap-r}\\
L^{(s)}_\theta\,&=& -\frac{1}{R^2\,r^2}\,\partial_\theta^2 - \frac{\cot\,\theta}{R^2\,r^2}\,\partial_\theta\label{eq:scalarlap-theta}\\
L^{(s)}_\phi\,&=& -\frac{1}{R^2\,r^2\,\sin^2\,\theta}\,\partial_\phi^2\label{eq:scalarlap-phi}
\eea
\end{subequations}
Though a covariant meaning only obtains when $L^{(s)}$ acts on a \emph{scalar}, in what follows we shall also view
it as a differential operator acting on any function of coordinates, not necessarily scalar functions.
Even though such a procedure may not have any manifestly coordinate-independent or global meaning, it is certainly meaningful 
locally as long as one works consistently within a particular coordinate system like, for example, the comoving coordinates.

\section{Killing vector equations}
The Killing vectors are solutions of eqn.(\ref{eq:killingeqns}).
Though these eqns are expressed in terms of the \emph{lower components} $\xi_\mu$, we shall present the Killing equations for both the 
upper components $\xi^\mu$ as well as the lower components. To obtain the former, one can write down Killing vector equations for the lower 
components 
and from them derive the corresponding upper component ones. Since the FLRW-metric is diagonal, this is fairly straightforward.
\subsection{Lower component KV eqns}
\be
\label{eq:kvttlower}
\xi_{t,t} = 0
\ee
\be
\label{eq:kvtrlower}
\xi_{t,r}+\xi_{r,t} -2\frac{{\dot R}}{R}\,\xi_r = 0
\ee
\be
\label{eq:kvtthetalower}
\xi_{t,\theta}+\xi_{\theta,t}-2\,\frac{{\dot R}}{R}\,\xi_\theta = 0
\ee
\be
\label{eq:kvtphilower}
\xi_{t,\phi}+\xi_{\phi,t} - 2\,\frac{{\dot R}}{R}\,\xi_\phi = 0
\ee
\be
\label{eq:kvrrlower}
\xi_{r,r}-\frac{kr}{1-kr^2}\,\xi_r-\frac{R{\dot R}}{1-kr^2}\,\xi_t =0
\ee
\be
\label{eq:kvrthetalower}
\xi_{r,\theta}+\xi_{\theta,r} -2\,\frac{\xi_\theta}{r}=0
\ee
\be
\label{eq:kvrphilower}
\xi_{r,\phi}+\xi_{\phi,r} - 2\frac{\xi_\phi}{r} = 0
\ee
\be
\label{eq:kvthetathetalower}
\xi_{\theta,\theta} -R{\dot R}r^2\,\xi_t+r(1-kr^2)\,\xi_r =0
\ee
\be
\label{eq:kvthetaphilower}
\xi_{\theta,\phi}+\xi_{\phi,\theta}-2\,\cot\theta\,\xi_\phi = 0
\ee
\be
\label{eq:kvphiphilower}
\xi_{\phi,\phi} - R{\dot R}\,r^2\,\sin^2\theta\,\xi_t+r(1-kr^2)\sin^2\theta\,\xi_r+\sin\theta\cos\theta\,\xi_\theta =0
\ee
\subsection{Upper component KV eqns}
Now we give the corresponding upper component equations.
\be
\label{eq:kvttupper}
\xi^t_{\,\,t}=0
\ee
\be
\label{eq:kvtrupper}
\frac{R^2}{1-kr^2}\,\xi^r_{,t}-\xi^t_{,r}=0
\ee
\be
\label{eq:kvtthetaupper}
\xi^t_{,\theta}-R^2\,r^2\,\xi^\theta_{,t} = 0
\ee
\be
\label{eq:kvtphiupper}
\xi^t_{,\phi}-R^2\,r^2\,\sin^2\theta\,\xi^\phi_{,t}=0
\ee
\be
\label{eq:kvrrupper}
\xi^r_{,r}+\frac{kr}{1-kr^2}\,\xi^r+\frac{{\dot R}}{R}\,\xi^t = 0
\ee
\be
\label{eq:kvrthetaupper}
\frac{\xi^r_{,\theta}}{r^2(1-kr^2)} + \xi^\theta_{,r} =0
\ee
\be
\label{eq:kvrphiupper}
\frac{\xi^r_{,\phi}}{r^2(1-kr^2)}+\sin^2\theta\xi^\phi_{,r} = 0
\ee
\be
\label{eq:kvthetathetaupper}
\xi^\theta_{,\theta}+\frac{\xi^r}{r}+\frac{{\dot R}}{R}\,\xi^t =0
\ee
\be
\label{eq:kvthetaphiupper}
\frac{\xi^\theta_{,\phi}}{\sin^2\theta}+\xi^\phi_{,\theta} = 0
\ee
\be
\label{eq:kvphiphiupper}
\xi^\phi_{,\phi}+\frac{{\dot R}}{R}\,\xi^t+\frac{\xi^r}{r}+\cot\theta\,\xi^\theta = 0
\ee
\section{Explicit solutions and their properties}
To motivate our work, we now write down the explicit solutions of these Killing equations for the generic case i.e for $R(t)$ not satisfying
the maximal symmetry condition of eqn.(\ref{eq:maximal}). Even though the four-manifold is not maximally symmetric, the $t=const$ - hypersurface
is still so.
\begin{subequations}
\label{eq:3dkvupper}
\bea
\xi^t &=&0\label{eq:3dkvupper-t}\\
\xi^r &=& \sqrt{1-kr^2}\left(\sin\theta (\cos\phi \delta a_x +\sin\phi \delta a_y) +\cos \theta \delta a_z\right)\label{eq:3dkvupper-r} \\
\xi^{\theta} &=& \frac{\sqrt{1-kr^2}}{r} \left[ \cos\theta (\cos\phi \delta a_x + \sin\phi \delta a_y) -\sin\theta \delta a_z \right] +\left( \sin\phi \delta b_x - \cos\phi \delta b_y \right)\label{eq:3dkvupper-theta} \\
\xi^{\phi} &=&\frac{\sqrt{1-kr^2}}{r} \left[ \frac{1}{\sin\theta} (\cos\phi \delta a_y - \sin\phi \delta a_x) \right] +\cot\theta \left( \cos\phi \delta b_x +\sin \phi \delta b_y \right)- \delta b_z\label{eq:3dkvupper-phi}
\eea
\end{subequations}
There are indeed 6 Killing vectors required by the maximal symmetry of the 3-manifold. Further restricting to the 2-sphere, consistency
requires $\xi^r=0$ which can be realised by setting ${\vec \delta a}=0$, one sees exactly three Killing vectors left, characterstic of a
maximally symmetric 2-manifold.

Focussing on $\xi^\phi$, one observes the occurrence of the following :
\be
\label{eq:phiexplicit}
\xi^\phi:\quad\quad 1\quad\quad \cot\theta\cos\phi\quad\quad \cot\theta\sin\phi \quad\quad \frac{\cos\phi}{\sin\theta} \quad\quad \frac{\sin\phi}{\sin\theta}
\ee
here 1 has been used to symbolise no dependence on the coordinates. Likewise, $\xi^\theta$ is seen to involve
\be
\label{eq:thetaexplicit}
\xi^\theta:\quad\quad \sin\theta \quad\quad \cos\theta\cos\phi \quad\quad \cos\theta\sin\phi \quad\quad \cos\phi \sin\phi
\ee
 
One notices here that $\frac{\xi^\theta}{\sin\theta}$ and $\xi^\phi$ involve the same five independent functions. Hence $\xi^\phi$
and $\frac{\xi^\theta}{\sin\theta}$ will obey the same differential equations. Before discussing them, let us introduce the differential
operators
\be
\label{eq:dthetaphi}
{\cal D}_\phi\equiv\,\partial_\phi\quad\quad {\cal D}_\theta \equiv\,\sin\theta\,\partial_\theta
\ee
These obviously differ from the covariant derivatives introduced in eqns.(\ref{eq:covderivs},\ref{eq:covderivcontra}).
The constant solution obeys
\be
\label{eq:constantmode}
{\cal D}_\phi 1 = 0 \quad\quad {\cal D}_\theta 1 = 0
\ee
The other four functions of eqn.(\ref{eq:phiexplicit}) are not eigenstates of the operators of eqn.{\ref{eq:dthetaphi}).
However, they are all eigenstates of ${\cal D}_\phi^2$ with eigenvalue -1. That is,
\be
\label{eq:dphisq}
{\cal D}_\phi^2\,\cos\phi = -\cos\phi \quad\quad {\cal D}_\phi^2\,\sin\phi = -\sin\phi
\ee

Let us examine the action of ${\cal D}_\theta$ on the $\theta$-dependence of the other four:
\bea
\label{eq:dthetaact}
{\cal D}_\theta\,\frac{1}{\sin\theta} =  -\frac{\cos\theta}{\sin\theta}\\
{\cal D}_\theta\,\frac{\cos\theta}{\sin\theta} = -\frac{1}{\sin\theta}
\eea
i.e they provide a doublet representation. It then follows that
\bea
\label{eq:dthetasq}
{\cal D}_\theta^2\,\frac{1}{\sin\theta} = \frac{1}{\sin\theta}\nonumber\\
{\cal D}_\theta^2\,\frac{\cos\theta}{\sin\theta} = \frac{\cos\theta}{\sin\theta}
\eea

Thus all the four functions have
\be
\label{eq:zeromodes}
{\cal D}_\phi^2 = -1\quad\quad {\cal D}_\theta^2=1
\ee
Therefore, all five functions of eqn.(\ref{eq:phiexplicit}) satisfy
\be
\label{eq:zeromodes2}
{\cal D}_\phi^2+{\cal D}_\theta^2 =0
\ee
In fact, eqn.(\ref{eq:zeromodes2}) follows directly from the Killing equations. To see this, note that
eqns.(\ref{eq:kvthetathetaupper},\ref{eq:kvphiphiupper}) can be combined into
\be
\label{eq:thetathetaphiphicombo}
\xi^\phi_{,\phi}\,-\xi^\theta_{,\theta}\,+\cot\theta\,\xi^\theta=0
\ee
and this can be recast as
\be
\label{eq:thetathetaphiphi2}
{\cal D}_\phi\,\xi^\phi-{\cal D}_\theta\,(\frac{\xi^\theta}{\sin\theta})=0
\ee
Likewise, eqn.(\ref{eq:kvthetaphiupper}) can be recast as
\be
\label{eq:thetaphiupper2}
{\cal D}_\phi\,\left(\frac{\xi^\theta}{\sin\theta}\right )+{\cal D}_\theta\,\xi^\phi=0
\ee
It is then straightforward to obtain eqn.(\ref{eq:zeromodes2}) by acting eqn.(\ref{eq:thetathetaphiphi2}) with ${\cal D}_\theta$,
and eqn.(\ref{eq:thetaphiupper2}) with ${\cal D}_\phi$.

On comparing with eqn.(\ref{eq:splitscalarlap}), one sees that all five functions satisfy
\be
\label{eq:scalarzero}
L^{(s)}_{\theta,\phi} = L^{(s)}_\theta+L^{(s)}_\phi = -\frac{{\cal D}_\theta^2+{\cal D}_\phi^2}{R^2\,r^2\,\sin^2\theta}
=0
\ee
To understand the nature of these zero modes, we make use of three important notions:\\
{\bf a) Are the functions well-defined?}\\
Except for the constant mode of eqn.(\ref{eq:phiexplicit}), the other modes are not defined at $\theta\,=\,0,\pi$. However, this
issue of whether functions are well-defined everywhere or not is rather subtle as the $\theta,\phi$  coordinates are not well
defined everywhere, and they break down precisely at $\theta\,=\,0,\pi$. The correct way of handling this is to introduce so called
\emph{atlases} on the two-sphere. We shall not go into those details because of their highly technical nature. 
It suffices to say that in some cases these are mere \emph{coordinate singularities} removable by choosing a different coordinate
system near the north and south poles. But this can not be done in all cases.\\

\noindent
{\bf b) Square integrability}\\
A function $f(\theta,\phi)$ is square-integrable if
\be
\label{eq:squareintegrability}
\int\,\sin\theta\,d\theta\,d\phi\: f(\theta,\phi)^2\, < \infty
\ee
As stated, this too depends on the choice of coordinates and the integral has to be made meaningful by choosing sufficient number of
atlases and transition functions. Just for the sake of classifying the solutions of eqn.(\ref{eq:scalarzero}), we shall evaluate the
integrals of eqn.(\ref{eq:squareintegrability}) as ordinary integrals.

\noindent
{\bf c) Normalizability}\\
Finally, If $\xi^\theta,\xi^\phi$ are the components of the vector $\xi$, the vector is normalizable if
\be
\label{eq:normalizability}
N = \int\,\sin\theta\,d\theta\,d\phi\:\left\{(\xi^\theta)^2\,+\,\sin^2\theta\,(\xi^\phi)^2\right\}\,<\,\infty
\ee
The three Killing vectors on the 2-sphere implied by eqns.(\ref{eq:3dkvupper}) are:
\be
\label{eq:2dKV}
\xi^\theta_{(1)}=0,\xi^\phi_{(1)}=1\quad\quad \xi^\theta_{(2)}=\sin\phi,\xi^\phi_{(2)}=\cot\theta\,\cos\phi\quad\quad \xi^\theta_{(3)}=-\cos\phi,
\xi^\phi_{(3)}=\cot\theta\sin\phi
\ee
Thus we see that $\xi_{(1)}$ corresponds to the constant zero mode; the scaled components are well defined everywhere, they are square integrable
and this Killing vector is normalizable. On the other hand, the scaled components of $\xi_{(2)}$ are neither well-defined everywhere nor square-integrable.
They are, what we call, \emph{irregular}. Nevertheless, this Killing vector is normalizable. The same holds for the Killing vector $\xi_{(3)}$.\\
 
It is also possible to construct the eigenstates of ${\cal D}_\theta$ itself. In this case, the eigenvalues will be
$\pm\,1$:
\be
\label{eq:dthetaeigen}
{\cal D}_\theta\:\left(\frac{1}{\sin\theta}\,\mp\,\frac{\cos\theta}{\sin\theta}\right ) = \pm\,
\left(\frac{1}{\sin\theta}\,\mp\,\frac{\cos\theta}{\sin\theta}\right ) 
\ee
If we denote these by $\chi_1^{\pm}$, it is clear that these are also eigenstates of ${\cal D}_\theta^2$ with the doubly 
degenerate eigenvalue 1. The eigenstates displayed in eqn.(\ref{eq:dthetasq}) are in fact linear combinations of these. 
We make this explicit by introducing
\be
\label{eq:S&C}
\chi_1^S = \frac{\chi_1^++\chi_1^-}{2}=\frac{1}{\sin\theta}\quad\quad \chi_1^C = \frac{\chi_1^--\chi_1^+}{2}=\frac{\cos\theta}{\sin\theta}
\ee
\subsection{The spectrum of ${\cal D}_\theta$}
We now complete the above discussion by working out all the zero modes of $L^{(s)}_{\theta,\phi}$. It is clear that single-valuedness requires
\be
\label{eq:phisinglevalue}
{\cal D}_\phi^2 = -n^2
\ee
with n an integer. Thus the zero modes in question must satisfy
\be
\label{eq:dthetasqgen}
{\cal D}_\theta^2 = n^2
\ee
This also means that these are superpositions of the eigenstates of ${\cal D}_\theta$ with eigenvalues $\pm\,n$. Now we construct the
eigenstates of ${\cal D}_\theta$. We restrict these to be finite sums of the form
\be
\label{eq:dthetaeigen}
\chi_n^{\pm}(\theta) = \sum\:\alpha_{p,q}^{\pm}\,\sin^p\theta\,\cos^q\theta
\ee
It is worth emphasising that eqn.(\ref{eq:dthetaeigen}) are not really restrictive. This form allows us to proceed with our analysis, at the
end of which we shall find \emph{unique} solutions to $\chi_n^{\pm}$. Let the maximum value of p be N and let the corresponding value of q be M. Let us consider the action of ${\cal D}_\theta$ on such a term:
\be
\label{eq:dthetaNMaction}
{\cal D}_\theta\:\sin^N\theta\cos^M\theta = N\,\sin^N\theta\,\cos^{M-1}\theta - (N+M)\,\sin^{N+2}\theta\,\cos^{M-1}\theta 
\ee
From this it follows that for the eigenstate of ${\cal D}_\theta$ to be a \emph{finite} sum of the type of eqn.(\ref{eq:dthetaeigen}),
N has to be \emph{negative} and M has to be \emph{positive}. If M is \emph{odd}, it can be reduced to one by using
\be
\label{eq:Mreduce}
\cos^{2M^\prime+1}\,\theta = \cos\theta\,(1-\sin^2\theta)^{M^\prime}
\ee
and rearranging. Likewise, if M is even, it can be reduced to zero. Thus we only need consider terms of the type
\be
\label{eq:newdthetaeigen}
\chi_n^{\pm}(\theta) = \sum\:(\alpha_{n,m}^{\pm}+\beta_{n,m}^{\pm}\,\cos\theta)\,\frac{1}{\sin^m\theta}\quad\quad
{\cal D}_\theta\,\chi_n^{\pm} = \pm\,n\,\chi_n^{\pm}(\theta)
\ee
Now we work out the generaliations of eqn.(\ref{eq:dthetaact}):
\bea
\label{eq:dthetaactgen}
{\cal D}_\theta\:\frac{1}{\sin^m\theta} = -m\,\frac{\cos\theta}{\sin^m\theta}\\
{\cal D}_\theta\:\frac{\cos\theta}{\sin^m\theta} = -m\,\frac{1}{\sin^m\theta}+(m-1)\,\frac{1}{\sin^{m-2}\theta}
\eea
From these, the generalizations of eqn.(\ref{eq:dthetaeigen}) follows:
\be
\label{eq:dthetaeigengen}
{\cal D}_\theta\,\left (\frac{1}{\sin^m\theta}\mp\frac{\cos\theta}{\sin^m\theta}\right ) = \pm\,m\,\left (\frac{1}{\sin^m\theta}\mp
\frac{\cos\theta}{\sin^m\theta}\right )\,\mp\,\frac{m-1}{\sin^{m-2}\theta}
\ee
The immediate consequence of this is that
\be
\label{eq:eigenfngen}
\alpha_{n,n}^{\pm} = \alpha\quad\quad \beta_{n,n}^{\pm} = \mp\,\alpha\quad\quad \alpha_{n,n-q}^{\pm}=\beta_{n,n-q}^{\pm}=0
\ee
Here ${\mathrm q}$ is any odd integer less than ${\mathrm n}$, and, $\alpha$ ia any non-zero constant that does not depend on n. Without loss
of generality it can be taken to be 1. On using eqn.(\ref{eq:dthetaeigengen}), the eigenvalue equation leads to the following recursion relation
for the coefficients $\alpha_{n,m}^{\pm},\beta_{n,m}^{\pm}$:
\bea
\label{eq:alphabetarecursion}
\pm\,n\,\alpha_{n,m}^{\pm} = (m+1)\,\beta_{n,m+2}^{\pm}\,-m\,\beta_{n,m}^{\pm}\\
\pm\,n\,\beta_{n,m}^{\pm} = -m\,\alpha_{n,m}^{\pm}
\eea
It is easy to solve these recursion relations:
\bea
\label{eq:alphabetanm}
\beta_{n,m}^{\pm} = -\frac{m(m+1)}{n^2-m^2}\,\beta_{n,m+2}^{\pm}\\
\alpha_{n,m}^{\pm} = \mp\,\frac{n}{m}\,\beta_{n,m}^{\pm}
\eea
It then follows that:
\bea
\label{eq:alphabetanm2}
\alpha_{n,m}^{\pm} = \alpha_{n,m}\\
\beta_{n,m}^{\pm}  = \mp\,\beta_{n,m}
\eea
In fact, eqn.(\ref{eq:alphabetanm}) imply that all the coefficients are uniquely fixed in terms of the constant $\alpha$ introduced in
eqn.(\ref{eq:eigenfngen}). In otherwords, the eigenvalues $\pm\,n$ of ${\cal D}_\theta$ are \emph{nondegenerate}. We cite some explicit
values:
\bea
\label{eq:alphabetaexplicit}
\beta_{n,n-2} = \frac{2-n}{4}\quad\quad \beta_{n,n-4} = \frac{(n-3)(n-4)}{32}\\
\alpha_{n,n-2} = -\frac{n}{4}\quad\quad \alpha_{n,n-4} = \frac{n(n-3)}{32}
\eea
Therefore, the eigenvalue $n^2$ of ${\cal D}_\theta^2$ is \emph{doubly degenerate}. The two independent eigenstates corresponding to
this are of course $\chi_n^{\pm}(\theta)$. Because of the degeneracy any linear combinations of these are also eigenstates with the same eigenvalue. Two particularly useful combinations are:
\bea
\label{eq:C&Seigens}
\chi_n^S = \frac{\chi_n^++\chi_n^-}{2} = \sum\:\frac{\alpha_{n,m}}{\sin^m\theta}\\
\chi_n^C = \frac{\chi_n^--\chi_n^+}{2} = \sum\:\frac{\beta_{n,m}\,\cos\theta}{\sin^m\theta}\\
\eea
It is immediately obvious that for $n\,\ge\,2$, none of the scaled components are either well-defined or square integrable, and, the 
corresponding vectors are not normalizable.
\section{Relation to Killing vectors}
Thus one sees that there are \emph{infinitely} many solutions to eqn.(\ref{eq:zeromodes2})! But the number of Killing vectors in the
present context can atmost be 10. Thus all but a few solutions are not admissible by the full set of Killing equations. In fact, solutions
with $n\,\ge\,2$ are excluded. We shall present the essence of the arguments leading to this conclusion, and skip the full details. From
the discussion so far, the possible contributions from $n\,\ge\,2$ to the most general solutions for $\frac{\xi^\theta}{\sin\theta}$ is:
\bea
\label{eq:genxitheta}
\frac{\xi^\theta}{\sin\theta} = \sum_{n\ge\,2}\:& &\cos\,n\phi\,\left (A_{n,+}(r,t)\chi_n^+(\theta)+A_{n,-}(r,t)\chi_n^-(\theta)\right )\nonumber\\
& &+\sin\,n\phi\left (B_{n,+}(r,t)\chi_n^+(\theta)+B_{n,-}(r,t)\chi_n^-(\theta)\right )
\eea
Eqns.(\ref{eq:thetathetaphiphi2},\ref{eq:thetaphiupper2}) and the fact that $\chi_n^{\pm}(\theta)$ are eigenfunctions of ${\cal D}_\theta$
with eigenvalues $\pm\,n$ then imply, for similar contributions to $\xi^\phi$:
\bea
\label{eq:genxiphi}
\xi^\phi = \sum_{n\ge\,2}\:& &\cos\,n\phi\,\left (-B_{n,+}(r,t)\chi_n^+(\theta)+B_{n,-}(r,t)\chi_n^-(\theta)\right )\nonumber\\
& &+\sin\,n\phi\left (A_{n,+}(r,t)\chi_n^+(\theta)-A_{n,-}(r,t)\chi_n^-(\theta)\right )
\eea
Let us apply eqn.(\ref{eq:kvtthetaupper}) to get
\bea
\label{eq:nge2}
\xi^t_{,\theta} = R^2(t)\,r^2\,\sin\theta\,
 \sum_{n\ge\,2}\:& &\cos\,n\phi\,\left ({A^\prime}_{n,+}(r,t)\chi_n^+(\theta)+{A^\prime}_{n,-}(r,t)\chi_n^-(\theta)\right )\nonumber\\
& &+\sin\,n\phi\left ({B^\prime}_{n,+}(r,t)\chi_n^+(\theta)+{B^\prime}_{n,-}(r,t)\chi_n^-(\theta)\right )
\eea
where $A^\prime$ denotes derivative wrt t etc. On introducing
\be
\label{eq:integralchi}
\eta_n^{\pm}(\theta) = \int^\theta\,d\theta^\prime\,\sin\theta^\prime\,\chi_n^{\pm}\,(\theta^\prime)
\ee
where constants of integration are already absorbed into $\eta_n^{\pm}$. We can integrate eqn.(\ref{eq:nge2}) to get(we only make explicit the $n\,\ge\,2$ terms):
\bea
\label{eq:nge3}
\xi^t = R^2(t)\,r^2\,
 \sum_{n\ge\,2}\:& &\cos\,n\phi\,\left ({A^\prime}_{n,+}(r,t)\eta_n^+(\theta)+{A^\prime}_{n,-}(r,t)\eta_n^-(\theta)\right )\nonumber\\
& &+\sin\,n\phi\left ({B^\prime}_{n,+}(r,t)\eta_n^+(\theta)+{B^\prime}_{n,-}(r,t)\eta_n^-(\theta)\right )
\eea
On combining this with eqn.(\ref{eq:kvtphiupper}), one gets
\bea
\label{eq:nge4}
&-&\,n\,R^2\,r^2\,\left\{\sin\,n\phi\,({A^\prime}_{n,+}\eta_n^++{A^\prime}_{n,-}\eta_n^-)-\cos\,n\phi\,({B^\prime}_{n,+}\eta_n^++{B^\prime}_{n,-}\eta_n^-)\right\}\nonumber\\
&=& R^2\,r^2\,\sin^2\theta\left\{\sin\,n\phi\,({A^\prime}_{n,+}\chi_n^+-{A^\prime}_{n,-}\chi_n^-)-\cos\,n\phi\,({B^\prime}_{n,+}\chi_n^+-{B^\prime}_{n,-}\chi_n^-)\right\}
\eea
Equating the coefficients of $\sin\,n\phi, \cos\,n\,\phi$, respectively, yields:
\bea
\label{eq:nge24}
-n\,[{A^\prime}_{n,+}\,\eta_n^+\,+\,{A^\prime}_{n,-}\,\eta^-] &=& \sin^2\theta\,[{A^\prime}_{n,+}\,\chi_n^+\,-\,{A^\prime}_{n,-}\,\chi_n^-]\\
-n\,[{B^\prime}_{n,+}\,\eta_n^+\,+\,{B^\prime}_{n,-}\,\eta^-] &=& \sin^2\theta\,[{B^\prime}_{n,+}\,\chi_n^+\,-\,{B^\prime}_{n,-}\,\chi_n^-]
\eea
We rewrite these in the forms:
\bea
\label{eq:nge25}
{A^\prime}_{n,+}\,(\sin^2\theta\,\chi_n^+\,+\,n\eta_n^+)&=& {A^\prime}_{n,-}\,(\sin^2\theta\,\chi_n^-\,-\,n\eta_n^-)\\
{B^\prime}_{n,+}\,(\sin^2\theta\,\chi_n^+\,+\,n\eta_n^+)&=& {B^\prime}_{n,-}\,(\sin^2\theta\,\chi_n^-\,-\,n\eta_n^-)
\eea
Before analysing the implications of these conditions, we derive an expression for what we call the \emph{leading} terms(denoted by the subscript 
L) of
$\eta_n^{\pm}$. By this we mean the terms with the \emph{most negative} powers of $\sin\theta$. For example,
\be
\label{eq:leadingchi}
\chi_{n,L}^{\pm} = \frac{1}{\sin^n\,\theta}\,\mp\,\frac{\cos\theta}{\sin^n\,\theta}
\ee
As will be clear shortly, it is enough to know $\eta_{n,L}^{\pm}$ for the proof of absence of $n\,\ge\,2$ terms. Towards this, we 
note the following two types of indefinite integrals, the first of which can be obtained trivially:
\be
\label{eq:typeIintegral}
I_m^{(1)} \equiv\,\int\,d\theta\,\frac{\cos\theta}{\sin^{m-1}\theta}\,= \frac{1}{2-m}\,\frac{\sin^2\theta}{\sin^m\,\theta}
\ee
It is to be noted that this integral for $m=2$ is equal to $\ln\,|\sin\,\theta|$, and therefore has a completely different 
$\theta$-dependence than the $\chi^{\pm}$. Further, $\eta_n^{\pm}$ for every even $n\,\ge\,2$ will have such terms.
The other integral
\be
\label{eq:typeIIintegral}
I_m^{(2)} \equiv\,\int\,d\theta\,\frac{1}{\sin^{m-1}\theta}
\ee
satisfies the recursion relation
\be
\label{eq:type2recursion}
I_m^{(2)} = \frac{m-3}{m-2}\,I_{m-2}^{(2)}\,-\,\frac{1}{m-2}\,\frac{\cos\theta}{\sin^m\theta}
\ee
The second type of integral, for $m=2$, equals $\ln\,|\tan\,\frac{\theta}{2}|$. This too has a $\theta$-dependence
completely different from the $\chi^{\pm}$ and again every even n eigenstate will contain such terms. Thus even values of 
${\mathrm n}$ with $n\,\ge\,2$ can be ruled out at once. We only
need to show their absence for odd values of n. But the proof of the latter is general enough to be applicable to all values of n.
It follows from eqn.(\ref{eq:type2recursion}) that
\be
\label{eq:type2leading}
I_{m,L}^{(2)} = \frac{1}{2-m}\,\frac{\cos\theta}{\sin^{m-2}\theta}
\ee
Putting everything together, we get
\be
\label{eq:etaL}
\eta_{n,L}^{\pm}(\theta) = \pm\,\frac{\sin^2\theta}{n-2}\,\chi_{n,L}^{\pm}(\theta)
\ee
Returning to eqns.(\ref{eq:nge25}), since the A's and B's do not mix, one can analyse them separately. Let us analyse the A-coefficients
first. The analysis of B-coefficients is identical. there are three cases to consider. 

{\bf Case I}: this is when none of the ${A^\prime}_n^{\pm}$ vanish. In that case we have
\be
\label{eq:nge2case1}
\frac{\sin^2\theta\,\chi_n^{-}\,-n\,\eta_n^{-}}{\sin^2\theta\,\chi_n^+\,+\,n\,\eta_n^{+}}
=\frac{{A^\prime}_{n,+}}{{A^\prime}_{n,-}}
\ee
Since the $A$ depend only on (r,t), and $\chi,\eta$ only on $\theta$, this is possible only if the two ratios of eqn.(\ref{eq:nge2case1}) 
are \emph{constants}, say, $K_n$. In particular
\be
\label{eq:nge2case1a}
\frac{\sin^2\theta\,\chi_n^{-}\,-n\,\eta_n^{-}}{\sin^2\theta\,\chi_n^+\,+\,n\,\eta_n^{+}} = K_n
\ee
On the other hand, it is seen that
\be
\label{eq:nge2case1b}
\frac{\sin^2\theta\,\chi_{n,L}^{-}\,-n\,\eta_{n,L}^{-}}{\sin^2\theta\,\chi_{n,L}^+\,+\,n\,\eta_{n,L}^{+}}
=\frac{1-\cos\theta}{1+\cos\theta}
\ee
Thus the ratio is not a constant even in leading order. Sub-leading terms having very different $\theta$-dependences, can not alter
this conclusion. Therefore both ${A^\prime}_{n,+}$ and ${A^\prime}_{n,-}$ can not be nonzero. 

{\bf Case II}: Let ${A^\prime}_{n,+}\ne 0$(the analysis
of the other possibility is identical). Then, from eqn.(\ref{eq:nge25}) it follows that
\be
\label{eq:nge2case2}
{A^\prime}_{n,-}(\sin^2\theta\,\chi_n^-\,-\,n\,\eta_n^-) = 0
\ee
But
\be
\label{eq:nge2case2a}
\sin^2\theta\,\chi_{n,L}^-\,-\,n\,\eta_{n,L}^- = \frac{2n-2}{n-2}\,\sin^2\theta\,\chi_{n,L}^-\,\ne\,0
\ee
means that even in leading order eqn.(\ref{eq:nge2case2}) implies ${A^\prime}_{n,-}=0$. In other words, all the A-coefficients must
vanish.

By a similar analysis one concludes that all the B-coefficients must also vanish. This completes the proof that zero modes with $n\,\ge\,2$
do not contribute to the Killing Vectors.
\section{Our conjecture}
Thus we have explicitly shown that for the $\theta,\phi$-submanifold, which in the present context is the 2-sphere and
hence maximally symmetric, the Killing vector component $\xi^\theta$ when scaled by $\frac{1}{\sin\theta}$, and the Killing vector component $\xi^\phi$
when trivially scaled, are zero modes of the scalar, not covariant, Laplacian on the sphere. On the basis of this, we make the following
conjecture:

{\bf Conjecture}:\emph{All components of the Killing vectors of FLRW-space in comoving coordinates, when scaled by 
suitable functions that depend on
which component is scaled, are zero modes of the corresponding scalar 
Laplacian. Some of these zero modes are irregular(in the sense that they are not well defined everywhere and/or not square-integrable).
This also holds for the maximally symmetric sub-manifolds with constant comoving coordinates. For example, the three dimensional $t= constant $ manifold.
}

Some comments are in order in this context. The operator that naturally acts on Killing vectors is the covariant Laplacian, which, in
the present context is also the so called vector-Laplacian. By natural action, we mean one which has coordinate independent significance, what some may refer to as geometrical significance.  The structure of this vector Laplacian, in all its generality, is pretty
complicated and it explicitly involves the knowledge of the Christoffel connection components (see, for example \cite{weinbook}). As will be explicitly shown in the next
section, the covariant Laplacian on Killing vectors does not vanish in general, vanishing only for Ricci-flat spaces. This particular demonstration can indeed be done in arbitrary coordinate systems. In what follows, all considerations of this paper hold only in comoving coordinate system.
In this coordinate system, it is rather remarkable and unexpected that by scaling the Killing vector components appropriately, one obtains zero modes of the scalar Laplacian!
We prove the conjecture by first deriving two alternate expressions for the covariant Laplacians, equating them, and finally, by a
systematic procedure to find the required scalings.

\section{Covariant Laplacian on Killing Vectors}
The object of this section is the evaluation of 
\be
\label{eq:kvbox}
\Box\,\xi^\mu = D^\nu\,\xi^\mu_{;\nu}
\ee
A straightforward evaluation using the definition of covariant derivatives, is of course possible. But it is tedious and requires 
the explicit values of the connection \cite{weinbook}. Here we propose two different ways of obtaining the desired result. Both of them exploit the fact
that the Killing vectors are solutions of the Killing vector equations, but do not need their explicit forms. The novelty of both
methods is that the connection components need not be used at all.
\subsection{Method based on commutator of covariant derivatives}
Recall that the commutator of covariant derivatives acting on any tensor field is proportional to both the tensor field and the
curvature tensor. Explicitly, eqn(6.5.2) of Weinberg \cite{weinbook} reads
\be
\label{eq:covdercommutator}
V^\lambda_{;\nu;\kappa} - V^\lambda_{;\kappa;\nu} = R^\lambda_{\sigma\nu\kappa}\,V^\sigma
\ee
Instead of eqn(\ref{eq:covdercommutator}),
we use eqn(6.5.1) of Weinberg:
\be
\label{eq:covdercommlower}
V_{\beta;\alpha;\gamma} - V_{\beta;\gamma;\alpha} = -R^\sigma_{\beta\alpha\gamma}\,V_\sigma
\ee
Applying to Killing vectors,
\be
\label{eq:kvcovdercommlower}
\xi_{\beta;\alpha;\gamma} - \xi_{\beta;\gamma;\alpha} = -R^\sigma_{\beta\alpha\gamma}\,\xi_\sigma
\ee
Hence
\be
\label{eq:kvbox2}
\Box\,\xi_\alpha= D^\beta\,\xi_{\alpha;\beta} = -D^\beta\,\xi_{\beta;\alpha} = -g^{\beta\gamma}\,\xi_{\beta;\alpha;\gamma}
\ee
Here the Killing equation $\xi_{\alpha;\beta}+\xi_{\beta;\alpha}=0$ was used. On using eqn(\ref{eq:kvcovdercommlower}), this becomes
\bea
\Box\,\xi_\alpha &=& -g^{\beta\gamma}\,\big\{\xi_{\beta;\gamma;\alpha} -\xi_\sigma\,R^\sigma_{\beta\alpha\gamma}\big\}\nonumber\\
&=&-(g^{\beta\gamma}\xi_{\beta;\gamma})_{;\alpha} +g^{\beta\gamma}\,R^\sigma_{\beta\alpha\gamma}\,\xi_\sigma\nonumber\\
&=& g^{\beta\gamma}\,R^\sigma_{\beta\gamma\alpha}\,\xi_\sigma
\eea
where we used that $g^{\beta\gamma}\,\xi_{\beta;\gamma}=0$ in accordance with the Killing equations.This can be further:  
\be
\label{eq:kvbox3}
\Box\,\xi_\alpha = R^{\sigma\gamma}_{\,\,\,\alpha\gamma}\,\xi_\sigma = R^{\gamma\sigma}_{\,\,\gamma\alpha}\,\xi_\sigma = R_{\sigma\alpha}\,\xi^\sigma
\ee
leading finally to
\be
\label{eq:kvbox4}
\Box\,\xi^\alpha = R^\alpha_{\,\,\sigma}\,\xi^\sigma
\ee
From eqn.(\ref{eq:kvbox4}) we see that the covariant Laplacian on Killing vectors is in general nonzero, and vanishes only for spaces
which are \emph{Ricci-flat}.
\subsection{The method of antisymmetric tensors}
The other method that allows the evaluation of covariant Laplacian is to note that because of the Killing equations, $\xi_{\mu;\nu}$
is actually an \emph{antisymmetric tensor} of rank two. One can make this explicit on rewriting
\be
\label{eq:antisymmetry}
2\xi_{\mu;\nu} = \xi_{\mu;\nu} - \xi{\nu;\mu}+\xi_{\mu;\nu}+\xi_{\nu;\mu}=\xi_{\mu;\nu}-\xi_{\nu;\mu}\equiv\,{\cal A}_{\mu\nu}=-{\cal A}_{\nu\mu}
\ee
This construction has double benefits; the first is that due to the symmetry
\be
\label{eq:connectionsymmetry}
\Gamma^\alpha_{\beta\gamma} = -\Gamma^\alpha_{\gamma\beta}
\ee
the tensor ${\cal A}_{\mu\nu}$ does not depend on the connection i.e
\be
\label{eq:connindep}
{\cal A}_{\mu\nu}\equiv \xi_{\mu;\nu}-\xi_{\nu;\mu} = \partial_\nu\,\xi_\mu - \partial_\mu\xi_\nu
\ee
The other benefit is that covariant derivatives of antisymmetric tensors also do not require the connections(see Weinberg \cite{weinbook} 4.7.10)!
\be
\label{eq:covdivantisym}
A^{\mu\nu}_{\,\,\mu} = \frac{1}{{\sqrt{|g|}}}\,\frac{\partial}{\partial x^\mu}\,(\sqrt{|g|}A^{\mu\nu})
\ee
where $g=det(g_{\mu\nu})$. Now we arrive at an alternate expression for $\Box\,\xi^\mu$:
\be
\label{eq:kvboxanti}
\Box\,\xi_\mu = D^\nu\,\xi_{\mu;\nu} = \frac{1}{2}\,D^\nu\,{\cal A}_{\mu\nu} 
\ee
For the upper components the result follows trivially from the covariant constancy of the metric $g_{\mu\nu;\alpha}=0$:
\be
\label{eq:kvboxupperanti}
\Box\,\xi^\mu = \frac{1}{2}\,{\cal A}^{\mu\nu}_{\,\,\nu} = \frac{1}{2\,\sqrt{|g|}}\,\partial_\nu\,\sqrt{|g|}\,{\cal A}^{\mu\nu}
\ee
Thus the use of connection is completely avoided.
\subsection{Evaluation of ${\cal A}^{\mu\nu}$}
In this subsection, we shall explicitly evaluate the six distinct ${\cal A}^{\mu\nu}$. The results are shown in three equivalent forms
that will be found useful later on. Only  for the first case of ${\cal A}^{rt}$ the intermediate steps in the evaluation will be given.
For the rest, only the results will be shown.

{\bf ${\cal A}^{rt}$}:
\begin{subequations}
\bea
\label{eq:A-rt}
{\cal A}^{rt} &=& g^{rr}g^{tt}(\xi_{r,t}-\xi_{t,r})\nonumber\\
&=& -\frac{1-kr^2}{R^2}\big\{(-\frac{R^2}{1-kr^2}\,\xi^r)_{,t} -\xi^t_{,r}\big\}\nonumber\\
&=& 2\,\frac{{\dot R}}{R}\,\xi^r +\xi^r_{,t}+\frac{1-kr^2}{R^2}\xi^t_{,r}\label{eq:A-rt-a}\\
&=& 2\,\frac{{\dot R}}{R}\,\xi^r +2\,\xi^r_{,t}\label{eq:A-rt-b}\\
&=& 2\,\frac{{\dot R}}{R}\,\xi^r+2\,\frac{1-kr^2}{R^2}\,\xi^t_{,r}\label{eq:A-rt-c}
\eea
\end{subequations}
In the last two steps eqn(\ref{eq:kvtrupper}) has been used to recast ${\cal A}^{rt}$ in two different, but simpler, ways.

{\bf ${\cal A}^{\theta t}$}:
\begin{subequations}
\bea
\label{eq:A-thetat}
{\cal A}^{\theta t} 
&=& 2\,\frac{{\dot R}}{R}\,+\xi^\theta_{,t}+\frac{\xi^t_{,\theta}}{R^2\,r^2}\label{eq:A-thetat-a}\\
&=& 2\,\frac{{\dot R}}{R}\,\xi^\theta + 2\xi^\theta_{,t}\label{eq:A-thetat-b}\\
&=& 2\,\frac{{\dot R}}{R}\,\xi^\theta + 2\,\frac{\xi^t_{,\theta}}{R^2\,r^2}\label{eq:A-thetat-c}
\eea
\end{subequations}
The last two steps were obtained by using the Killing eqn(\ref{eq:kvtthetaupper}).

{\bf ${\cal A}^{\phi t}$}:
\begin{subequations}
\bea
\label{eq:A-phit}
{\cal A}^{\phi t}
&=& 2\,\frac{{\dot R}}{R}\,\xi^\phi + \xi^\phi_{,t}+\frac{1}{R^2\,r^2\,\sin^2\theta}\,\xi^t_{,\phi}\label{eq:A-phit-a}\\
&=& 2\,\frac{{\dot R}}{R}\,\xi^\phi + 2\,\xi^\phi_{,t}\label{eq:A-phit-b}\\
&=& 2\,\frac{{\dot R}}{R}\,\xi^\phi+\frac{2}{R^2\,r^2\,sin^2\theta}\,\xi^t_{,\phi}\label{eq:A-phit-c}
\eea
\end{subequations}
Killing eqn(\ref{eq:kvtphiupper}) was used.

{\bf ${\cal A}^{r\theta}$}:
\begin{subequations}
\bea
\label{eq:A-rtheta}
{\cal A}^{r\theta}
&=& 2\,\frac{1-kr^2}{R^2\,r}\,\xi^\theta\,-\frac {1}{R^2\,r^2}\,\xi^r_{,\theta}\,+\frac{1-kr^2}{R^2}\,\xi^\theta_{,r}\label{A-rtheta-a}\\
&=& 2\,\frac{1-kr^2}{R^2\,r}\,\xi^\theta\,-2\,\frac {1}{R^2\,r^2}\,\xi^r_{,\theta}\,\label{A-rtheta-b}\\
&=& 2\,\frac{1-kr^2}{R^2\,r}\,\xi^\theta\,+2\,\frac{1-kr^2}{R^2}\,\xi^\theta_{,r}\label{eq:A-rtheta-c}
\eea
\end{subequations}
Killing eqn(\ref{eq:kvrthetaupper}) was used.

{\bf ${\cal A}^{r\phi}$}:
\begin{subequations}
\bea
\label{eq:A-rphi}
{\cal A}^{r\phi} 
&=& \frac{2(1-kr^2)}{R^2\,r}\,\xi^\phi +\frac{1-kr^2}{R^2}\,\xi^\phi_{,r} - \frac{1}{R^2\,r^2\sin^2\theta}\,\xi^r_{,\phi}\label{eq:A-rphi-a}\\
&=& \frac{2(1-kr^2)}{R^2\,r}\,\xi^\phi +2\,\frac{1-kr^2}{R^2}\,\xi^\phi_{,r} \label{eq:A-rphi-b}\\
&=& \frac{2(1-kr^2)}{R^2\,r}\,\xi^\phi  - 2\,\frac{1}{R^2\,r^2\sin^2\theta}\,\xi^r_{,\phi}\label{eq:A-rphi-c}
\eea
\end{subequations}
Killing eqn(\ref{eq:kvrphiupper}) was used.

{\bf ${\cal A}^{\phi\theta}$}:
\begin{subequations}
\bea
\label{eq:A-phitheta}
{\cal A}^{\phi\theta} 
&=& -\frac{2}{R^2\,r^2}\,\cot\theta\,\xi^\phi-\frac{1}{R^2\,r^2}\,\xi^\phi_{,\theta}+\frac{1}{R^2\,r^2\,\sin^2\theta}\,\xi^\theta_{,\phi}\label{eq:A-phitheta-a}\\
&=& -\frac{2}{R^2\,r^2}\,\cot\theta\,\xi^\phi-\frac{2}{R^2\,r^2}\,\xi^\phi_{,\theta}\label{eq:A-phitheta-b}\\
&=& -\frac{2}{R^2\,r^2}\,\cot\theta\,\xi^\phi+\frac{2}{R^2\,r^2\,\sin^2\theta}\,\xi^\theta_{,\phi}\label{eq:A-phitheta-c}
\eea
\end{subequations}
Killing eqn(\ref{eq:kvthetaphiupper}) was used.
\subsection{Evaluations of $\Box \xi^\mu$ using ${\cal A}^{\mu\nu}$.}
The method based on commutation of covariant
derivatives simply gives $R^\mu_{\,\,\sigma}\,\xi^\sigma$ and as the mixed component Ricci tensor is diagonal, one simply gets a multiple
of the respective upper component Killing vector. We now describe the antisymmetric tensor method. 

From eqn(\ref{eq:kvboxupperanti})
\be
\label{eq:boxupperanti-2}
\Box \xi^\mu = \frac{1}{2\sqrt{|g|}}\,\frac{\partial}{\partial x^\nu}\,\sqrt{|g|}\,{\cal A}^{\mu\nu}
\ee
We show the details and various nuances only for $\Box\,\xi^\theta$. For the rest, we simply quote the final results.

{\bf (i) $\Box \xi^ \theta$}:

We calculate the contribution of $\nu=t$ in 
two ways to clarify some issues.
the contribution of the $\nu = t$ term is:
\bea
\label{eq:boxtheta1a}
\Box \xi^\theta|_{1a} &=& \frac{1}{2\sqrt{|g|}}\,(\sqrt{|g|}\,{\cal A}^{\theta t})_{,t}\nonumber\\
&=&\frac{1}{2}\,\left[3\,\frac{{\dot R}}{R}\,{\cal A}^{\theta t}+\partial_t\,{\cal A}^{\theta t}\right]\nonumber\\
&=&\frac{1}{2}\,\left[3\,\frac{{\dot R}}{R}\,\left\{2\,\frac{{\dot R}}{R}\,\xi^\theta\,+R^{-2}\,r^{-2}\,\xi^t_{,\,\theta}+\xi^\theta_{,\,t}\right\}\right]\nonumber\\
&+&\frac{1}{2}\,\left[\left\{2(\frac{{\ddot R}}{R}-\frac{{\dot R}^2}{R^2})\xi^\theta+2\,\frac{{\dot R}}{R}\,\xi^\theta_{,\,t}-2R^{-3}\,{\dot R}\,r^{-2}\,\xi^t_{,\,\theta}+
R^{-2}r^{-2}(\xi^t_{,\,t})_{,\,\theta}+\xi^\theta_{,\,t,\,t}\right\}\right]\nonumber\\
&=&\frac{1}{2}\,\left[\left(\frac{4{\dot R}^2+2\,R\,{\ddot R}}{R^2}\right)\,\xi^\theta+\,6\,\frac{{\dot R}}{R}\,\xi^\theta_{\,,t}
+\partial_t^2\,\xi^\theta\,\right]
\eea
In arriving at eqn.(\ref{eq:boxtheta}), we used the form of eqn.(\ref{eq:A-thetat-a}) for ${\cal A}^{\theta t}$. The following 
Killing equations (eqns.(\ref{eq:kvttupper},\ref{eq:kvtthetaupper})) were used in simplifying the final expression:
\be
\xi^t_{\,,t}=0\quad\quad \xi^t_{\,,\theta} = R^2\,r^2\,\xi^\theta_{\,,t}
\ee
\subsubsection{Alternate evaluation}
Now we derive eqn(\ref{eq:boxtheta1a}) by using the simplified expression of eqn.(\ref{eq:A-thetat-b}) we had obtained for 
${\cal A}^{\theta t}$:
\bea
\label{eq:boxtheta2a}
\Box \xi^\theta|_{2a} &=& \frac{1}{2\sqrt{|g|}}\,(\sqrt{|g|}\,{\cal A}^{\theta t})_{,t}\nonumber\\
&=&\frac{1}{2}\,\left[3\,\frac{{\dot R}}{R}\,{\cal A}^{\theta t}+\partial_t\,{\cal A}^{\theta t}\right]\nonumber\\
&=& \frac{1}{2}\,\left[3\frac{{\dot R}}{R}\,\left\{2\,\frac{{\dot R}}{R}\,\xi^\theta+2\,\xi^\theta_{\,,t}\right\}
+\left\{2\,\left(\frac{{\ddot R}}{R}-\frac{{\dot R}^2}{R^2}\right)\,\xi^\theta+2\,\frac{{\dot R}}{R}\,\xi^\theta_{\,,t}+2\,\partial_t^2\,\xi^\theta\,\right\}\right]\nonumber\\
&=&\,\frac{1}{2}\,\left[\frac{2\,R\,{\ddot R}+4\,{\dot R}^2}{R^2}\,\xi^\theta+8\,\frac{{\dot R}}{R}\,\xi^\theta_{\,,t}+2\partial_t^2\,\xi^\theta\right]
\eea
\subsubsection{Comparison}
A comparison of the two evaluations i.e eqn(\ref{eq:boxtheta1a}) and eqn(\ref{eq:boxtheta2a}) reveals the difference:
\be
\label{eq:boxthetadiff}
\partial_t^2\,\xi^\theta+2\,\frac{{\dot R}}{R}\,\xi^\theta_{\,,t}\equiv \frac{1}{R^2}\,\partial_t\,\left(R^2\,\xi^\theta_{\,,t}\right )
\ee
For the two forms to be consistent, one requires 
\be
\label{eq:boxthetaconsistency}
\partial_t\,\left(R^2\,\xi^\theta_{\,,t}\right )=0
\ee
Indeed, on differentiating eqn.(\ref{eq:kvtthetaupper}) wrt time and using eqn.(\ref{eq:kvttupper}), one sees that this consistency 
condition is satisfied.
The lesson is that different evaluations of $\Box$ may turn up different forms. They can all be mutualy consistent if their differences
vanish on the solutions of the Killing equations.

Now we evaluate the contribution from $\nu=r$. From now on we shall only use the simplified forms of ${\cal A}^{\mu\nu}$.
We skip the intermediate steps and only give the final result.
\be
\label{eq:boxthetab}
\Box \xi^\theta|_{b} 
= -\frac{1}{2}\,\left[\frac{2-4kr^2}{R^2\,r^2}\,\xi^\theta+\frac{6-8kr^2}{R^2\,r}\,\xi^\theta_{\,,r}+2\,\frac{1-kr^2}{R^2}\,\partial_r^2\,\xi^\theta\right]
\ee
This fully agrees with the expression obtained by using the unsimplified antisymmetric tensor.

Finally, we evaluate the contribution of $\nu = \phi$. Since the metric has no explicit $\phi$ dependence, this is really straihtforward:
\be
\label{eq:boxthetac}
\Box \xi^\theta|_{c} 
= \frac{1}{2}\,\left[\frac{2\,\cot\theta}{R^2\,r^2}\,\xi^\phi_{\,,\phi}-\frac{2}{R^2\,r^2\,\sin^2\theta}\,\partial_\phi^2\,\xi^\theta\right]
\ee
The unsatisfactory feature here is that $\Box\xi^\theta$ is involving $\xi^\phi$. We remedy this by eliminatig $\xi^\phi$ upon using the
Killing equations (\ref{eq:kvthetathetaupper},\ref{eq:kvphiphiupper}):
\be
\label{eq:xiphielim}
\xi^\phi_{\,,\phi}-\xi^\theta_{\,,\theta}+\cot\theta\,\xi^\theta = 0
\ee
Going back to eqn(\ref{eq:boxthetac}),
\bea
\label{eq:xiphielim2}
2\,\frac{\cot\theta}{R^2\,r^2}\,\xi^\phi_{\,,\phi}
&=&\frac{2}{R^2\,r^2}\,\cot\theta\,\left(\xi^\theta_{\,,\theta}-\cot\theta\,\xi^\theta\right)\nonumber\\
&=&\frac{2}{R^2\,r^2}\cot\theta\,\xi^\theta_{\,,\theta}-\frac{2}{R^2\,r^2\,sin^2\theta}\,\xi^\theta+\frac{2}{R^2\,r^2}\,\xi^\theta
\eea
Putting everthing together, the final reslt for $\Box \xi^\theta$ is
\bea
\label{eq:boxtheta}
\Box\,\xi^\theta &=& \frac{1}{2}\,\left[\left(\frac{4{\dot R}^2+2\,R\,{\ddot R}+4\,k}{R^2}\right)\,\xi^\theta+2\,\partial_t^2\,\xi^\theta+8\,\frac{{\dot R}}{R}\,\partial_t\,\xi^\theta\right]\nonumber\\
&-&\frac{1}{2}\,\left[\frac{6-8kr^2}{R^2\,r}\,\xi^\theta_{\,,r}+2\,\frac{1-kr^2}{R^2}\,\partial_r^2\,\xi^\theta\right]\nonumber\\
&-&\frac{1}{2}\,\left[2\,\frac{1}{R^2\,r^2\,\sin^2\theta}\,\partial_\phi^2\,\xi^\theta-2\,\frac{\cot\theta}{R^2\,r^2}\xi^\theta_{\,,\theta}+\frac{2}{R^2\,r^2\sin^2\theta}\,\xi^\theta\right]
\eea
It is worthwhile to contrast this with the expression one would have obtained by using the unsimplified expressions for the antisymetric
tensors:
\bea
\label{eq:boxthetaold}
\Box\,\xi^\theta &=& \frac{1}{2}\,\left[\left(\frac{4{\dot R}^2+2\,R\,{\ddot R}+4\,k}{R^2}\,\right)\,\xi^\theta+\partial_t^2\xi^\theta
+6\,\frac{{\dot R}}{R}\,\partial_t\,\xi^\theta\right]\nonumber\\
&-&\frac{1}{2}\,\left[\frac{6-8\,kr^2}{R^2\,r}\xi^\theta_{\,,r}+\frac{2(1-kr^2)}{R^2}\,\partial_r^2\,\xi^\theta\right]\nonumber\\
&-&\frac{1}{2}\,\left[\frac{1}{R^2\,r^2\,\sin^2\theta}\,\partial_\phi^2\,\xi^\theta -\frac{\cot\theta}{R^2\,r^2}\,\partial_\theta\,\xi^\theta+
\frac{1}{R^2\,r^2\,\sin^2\theta}\,\xi^\theta-\frac{2}{R^2\,r^2}\,\partial_\theta^2\,\xi^\theta\right]
\eea
\subsubsection{Issue of second derivatives}
Examination of these two forms of $\Box\,\xi^\theta$ reveals that while the former has no $\partial_\theta^2\,\xi^\theta$ terms at all,
the latter has, but with a 'wrong' sign from what may be naively expected from a Laplacian. The coefficients of other second-derivatives
also change from one form to another. As already remarked, as long as the differences vanish on solutions of Killing equations, there is
nothing to worry. We can do better if we can relate the second derivatives to lower derivatives. This is explicitly demonstrated now. Let
us start with the eqn(\ref{eq:kvthetathetaupper}):
\be
\xi^\theta_{\,,\theta}+\frac{\xi^r}{r}+\frac{{\dot R}}{R}\,\xi^t = 0
\ee
Differentiating this wrt $\theta$,
\be
\label{eq:dbltheta1}
\partial_\theta^2\,\xi^\theta+\frac{1}{r}\,\xi^r_{\,,\theta}+\frac{{\dot R}}{R}\,\xi^t_{\,,\theta}=0
\ee
Upon using the Killing eqn(\ref{eq:kvtthetaupper}) and eqn(\ref{eq:kvrthetaupper}), we get
\be
\label{eq:dbltheta2}
\partial_\theta^2\,\xi^\theta - r(1-kr^2)\,\partial_r\,\xi^\theta+R{\dot R}\,r^2\,\partial_t\,\xi^\theta =0
\ee
Therefore, suitable multiples of eqn(\ref{eq:dbltheta2}) can be added to get the theta-double derivatives of the desired type. Actually our eqn(\ref{eq:boxtheta}) has all other double derivatives of right sign and coefficients, so adding the right multiple of double-theta derivative
eqn can restore all the double derivative structures of the scalar Laplacian.

Indeed, upon subtracting $R^2$ times eqn.(\ref{eq:dbltheta2}) from eqn.(\ref{eq:boxtheta}), one gets,
\begin{eqnarray}
\label{eq:boxthetafin} 
\Box \xi^{\theta} &=& \left\lbrace \frac{2(\dot{R(t)})^2 + \ddot{R(t)} R(t) + 2k}{R^2(t)} \right\rbrace \xi^{\theta}\nonumber\\
 &+& \partial^2_t \xi^{\theta} - R^{-2}(t) (1-kr^2) \partial^2_r \xi^{\theta} - R^{-2}(t) r^{-2} \partial^2_{\theta} \xi^{\theta} 
- R^{-2}(t) r^{-2} \sin^{-2}\theta \partial^2_{\phi} \xi^{\theta}\nonumber\\
&+& 3\frac{\dot{R(t)}}{R(t)} \partial_t  \xi^{\theta} - R^{-2}(t) \left[ \frac{2(1-kr^2)}{r} -kr  \right] \partial_r \xi^{\theta} 
+R^{-2}(t) r^{-2} \cot\theta \partial_{\theta} \xi^{\theta} -R^{-2}(t) r^{-2} \sin^{-2}\theta \xi^{\theta}\nonumber\\
\end{eqnarray}
{\bf (ii) $\Box\,\xi^t$}:

All the nuances we encountered in the evaluation of $\Box\,\xi^\theta$ are present for all other Killing vectors also. We skip all such details,
and simply present the final result:

\begin{eqnarray}
\label{eq:boxxit} 
\Box \xi^{t}&=& 3\left(\frac{\dot{R(t)}}{R(t)}\right)^2 \xi^t + \partial^2_t \xi^t - R^{-2}(t) (1-kr^2) \partial^2_r \xi^t 
- R^{-2}(t) r^{-2} \partial^2_{\theta} \xi^t - R^{-2}(t) r^{-2} \sin^{-2}\theta \partial^2_{\phi} \xi^{t}\nonumber\\
&-&R^{-2}(t) \left[ \frac{2(1-kr^2)}{r} - kr  \right] \partial_r \xi^t - R^{-2}(t) r^{-2} \cot\theta \partial_{\theta} \xi^t 
\end{eqnarray}
{\bf (iii) $\Box\,\xi^r$}:

The final expression is:
\begin{eqnarray}
\label{eq:boxxir} 
\Box \xi^{r}&=& 
\partial^2_t \xi^r - R^{-2}(t)(1-kr^2) \partial^2_r \xi^r - R^{-2}(t) r^{-2} \partial^2_{\theta} \xi^r 
- R^{-2}(t) r^{-2} \sin^{-2}\theta \partial^2_{\phi} \xi^r \nonumber\\
&+& 3 \frac{\dot{R(t)}}{R(t)} \partial_t \xi^r + R^{-2}(t) \left[ \frac{2(1-kr^2)}{r} - kr \right] \partial_r \xi^r 
- R^{-2}(t) r^{-2} \cot\theta \partial_{\theta} \xi^r \nonumber\\
&-& R^{-2}(t) \left[ \frac{2(1-kr^2)}{r^2} +k + 2 \frac{(kr)^2}{(1-kr^2)} \right] \xi^r 
+\left\lbrace  \frac{2(\dot{R(t)})^2 + \ddot{R(t)} R(t) + 2k}{R^2(t)} \right\rbrace  \xi^r \nonumber\\
\end{eqnarray}
And finally, the case of $\Box \xi^\phi$}:

{\bf (iv) $\Box \xi^\phi$}:

\begin{eqnarray}
\label{eq:boxxiphi} 
\Box \xi^{\phi} &=& 
  \partial^2_t \xi^{\phi} - R^{-2}(t) (1-kr^2) \partial^2_r \xi^{\phi} - R^{-2}(t) r^{-2} \partial^2_{\theta} \xi^{\phi} 
- R^{-2}(t) r^{-2} \sin^{-2}\theta \partial^2_{\phi} \xi^{\phi}\nonumber\\
&+& 3\frac{\dot{R(t)}}{R(t)} \partial_t  \xi^{\phi} - R^{-2}(t) \left[ \frac{2(1-kr^2)}{r} - kr \right] \partial_r \xi^{\phi} 
- R^{-2}(t) r^{-2} \cot\theta \partial_{\theta} \xi^{\phi} \nonumber\\
&+& \left\lbrace \frac{2(\dot{R(t)})^2 + \ddot{R(t)} R(t) + 2k}{R^2(t)} \right\rbrace \xi^{\phi}
\end{eqnarray}

\section{Proof of the conjecture}
The proof makes use of the eqns.(\ref{eq:mixedriccivalues},\ref{eq:kvbox4},\ref{eq:scalarlap},\ref{eq:scalarlapsplit},\ref{eq:splitscalarlap}).

{\bf (a) $\xi^\phi$}:

Let us consider the case of $\xi^\phi$ first. In addition to the abovementioned equations, we also use eqn.(\ref{eq:boxxiphi}). Then it
follows that
\be
\label{eq:phiconjecture}
0 = \Box\,\xi^\phi - R^\phi_{\,\,\phi}\,\xi^\phi = L^{(s)}\,\xi^\phi
\ee
Thus we have shown that $\xi^\phi$, without any scaling, or, with trivial scaling, is a zero mode of the scalar Laplacian.

{\bf (b) $\xi^t$}:

Next, we analyse the $\xi^t$ case. Using eqn.(\ref{eq:boxxit}), we find that
\be
\label{eq:tconjecture}
0=\Box\,\xi^t\,-\,R^t_{\,\,t}\,\xi^t = (L^{(s)}_r+L^{(s)}_\theta+L^{(s)}_\phi)\,\xi^t+
\left\{\partial_t^2\,\xi^t+(3\frac{{\dot R}^2}{R^2}-3\frac{{\ddot R}}{R})\xi^t\right\}
\ee
where we have also made use of eqns.(\ref{eq:scalarlap-r},\ref{eq:scalarlap-theta}, \ref{eq:scalarlap-phi}).
Unlike the case of $\xi^\phi$, the rhs of eqn.(\ref{eq:tconjecture}) does not equal $L^{(s)}\,\xi^t$. For that to have happened, the terms
within $\left\{\ldots\right\}$, called ${\cal L}_t\xi^t$ below, should have equalled $L^{(s)}_t\,\xi^t$:
\be
\label{eq:calLt}
{\cal L}_t\,\xi^t = 
\partial_t^2\,\xi^t+(3\frac{{\dot R}^2}{R^2}-3\frac{{\ddot R}}{R})\xi^t
\ee
If a function $f_t(t)$ could be found such that
\be
\label{eq:ftscaling}
f_t(t)\,L^{(s)}_t\,\frac{\xi^t}{f_t(t)}={\cal L}_t\,\xi^t,
\ee
then it would follow that
\be
\label{eq:tconjecture2}
0 = \Box\,\xi^t\,-\,R^t_{\,\,t}\,\xi^t = f_t(t)\,L^{(s)}\,\frac{\xi^t}{f_t(t)}\rightarrow\,L^{(s)}\,\frac{\xi^t}{f_t(t)}=0
\ee
It is worth pointing out at this stage that although the scalar Laplacian $L^{(s)}$ is supposed to act only on scalars, and so also each
component of the split introduced in eqn.(\ref{eq:splitscalarlap}), we have interpreted them as differential operators acting on any function, 
not necessarily scalars. The basis for the assertion of eqn.(\ref{eq:tconjecture2}) is that scaling $\xi^t$ by a function that only depends 
on t will satisfy
\be
\label{eq:ftscalingfixed}
f_t(t)\,L^{(s)}_{r,\theta,\phi}\,\frac{\xi^t}{f_t(t)}= L^{(s)}_{r,\theta,\phi}\,\xi^t,
\ee

Thus the crux of proving our conjecture for $\xi^t$ is to find a $f_t(t)$ satisfying the eqn.(\ref{eq:ftscaling}). Using the explicit
form of $L^{(s)}_t$ as given in eqn.(\ref{eq:scalarlap-t}), this equation can be cast as:
\be
\label{eq:ftscaling2}
\partial_t^2\,\xi^t+(3\,\frac{{\dot R}}{R}\,-2\,\frac{f_t^\prime}{f_t}\,)\,\partial_t\,\xi^t\,+(-3\,\frac{{\dot R}}{R}\,\frac{f_t^\prime}{f_t}
-\frac{{f_t}^{\prime\prime}}{f_t}\,+2\,\frac{{f_t^\prime}^2}{f_t^2})\,\xi^t = \partial_t^2\,\xi^t+3\,(\frac{{\dot R}^2}{R^2}-\frac{{\ddot R}}{R})\,\xi^t
\ee
In what follows, $f^\prime$ shall stand for the derivative of f wrt its argument. For a general function $\xi^t$, no $f_t(t)$ can be found satisfying eqn.(\ref{eq:ftscaling2})! This is because 
it is not possible to simultaneously match the coefficients of $\partial_t\,\xi^t$ and $\xi^t$. However, $\xi^t$ being a Killing vector, satisfies eqn.(\ref{eq:kvttupper}) and eqn.(\ref{eq:ftscaling2}) reduces to
\be
\label{eq:ftscaling4}
(-3\,\frac{{\dot R}}{R}\,\frac{f_t^\prime}{f_t}
-\frac{{f_t}^{\prime\prime}}{f_t}\,+2\,\frac{{f_t^\prime}^2}{f_t^2})\,= 3\,(\frac{{\dot R}^2}{R^2}-\frac{{\ddot R}}{R})\,
\ee
whenever $\xi^t\,\ne\,0$. The solution of this equation is
\be
\label{eq:ftscaling3}
f_t(t) = \lambda\,R^3(t)
\ee
where $\lambda$ is a constant, amounting to trivial scaling, and can be set to unity without any loss of generality. Therefore, we have the
result
\be
\label{eq:ftscalingfin}
L^{(s)}\,\left (\frac{\xi^t}{R^3(t)}\right )\,=\,0
\ee

Though in this case, the issue of matching the coefficient of the first derivative of the Killing vector wrt 
the appropriate variable i.e $\partial_t\,\xi^t$ resolves itself trivially because of the Killing equations, 
we have made a pointed reference to it because in the cases of $\xi^r,\xi^\phi$ such terms are there, and the 
task of finding a suitable scaling is rather constrained. It is by no means obvious that a \emph{single} scaling function can
match the coefficients of both $\xi, \partial\,\xi$ for the analogs of eqn.(\ref{eq:ftscaling}) for all $\xi$.

{\bf (c) $\xi^\theta$}:

Turning to our result for $\Box\,\theta$ in eqn.(\ref{eq:boxthetafin}) we see
\be
\label{eq:thetaconjecture}
0 = \Box\,\xi^\theta\,-\,R^\theta_{\,\,\theta}\,\xi^\theta = (L^{(s)}_t+L^{(s)}_r+L^{(s)}_\phi)\,\xi^\theta+{\cal L}_\theta\,\xi^\theta
\ee
with
\be
\label{eq:calLtheta}
{\cal L}_\theta\,\xi^\theta = -\frac{1}{R^2\,r^2}\,\partial_\theta^2\,\xi^\theta\,+\,\frac{\cot\theta}{R^2\,r^2}\,\partial_\theta\,\xi^\theta\,
-\,\frac{1}{R^2\,r^2\,\sin^2\theta}\,\xi^\theta
\ee
Thus we see that ${\cal L}_\theta\,$ differes from $L^{(s)}_\theta$ in both $\partial_\theta\xi^\theta$ and $\xi^\theta$ terms. While the 
former is with the wrong sign, the latter is totally extra. As before, one seeks a scaling function $f_\theta(\theta)$ that only depends on
$theta$ so that 
\be
\label{eq:fthetascaling}
f_\theta(\theta)\,L^{(s)}_\theta\,\frac{\xi^\theta}{f_\theta(\theta)} = {\cal L}_\theta\,\xi^\theta\quad\quad
f_\theta(\theta)\,L^{(s)}_{t,r,\phi}\,\frac{\xi^\theta}{f_\theta(\theta)} =  L^{(s)}_{t,r,\phi}\,\xi^\theta
\ee
Using the explicit form of $L^{(s)}_\theta$ from eqn.(\ref{eq:scalarlap-theta}), we can expand this equation to get:
\bea
\label{eq:fthetascaling}
& &-\frac{1}{R^2\,r^2}\,\partial_\theta^2\,\xi^\theta\,+\,\left(\frac{2}{R^2\,r^2}\,\frac{f_\theta^\prime}{f_\theta}-\frac{\cot\theta}{R^2\,r^2}\right)\,\partial_\theta\,\xi^\theta
+\left\{\frac{\cot\theta}{R^2\,r^2}\,\frac{f_\theta^\prime}{f_\theta}+\frac{1}{R^2\,r^2}\,\left(\frac{f_\theta^{\prime\prime}}{f_\theta}-
\frac{2\,{f_\theta^\prime}^2}{f_\theta^2}\right)\right\}\,\xi^\theta\nonumber\\
&=&-\frac{1}{R^2\,r^2}\,\partial_\theta^2\,\xi^\theta\,+\,\frac{\cot\theta}{R^2\,r^2}\,\partial_\theta\,\xi^\theta\,-\,\frac{1}{R^2\,r^2\,\sin^2\theta}\,\xi^\theta
\eea
Matching the coefficients of $\partial_\theta\,\xi^\theta$ on both sides of eqn.(\ref{eq:fthetascaling}),
\be
\label{eq:fthetascaling2}
\frac{2}{R^2\,r^2}\,\frac{f_\theta^\prime}{f_\theta} -\frac{\cot\theta}{R^2\,r^2}\quad\rightarrow\quad \frac{f_\theta^\prime}{f_\theta}
=\cot\theta\quad\,f_\theta = \sin\theta
\ee
The consistency condition for the solution $f_\theta=\sin\theta$ is that the coefficients of $\xi^\theta$ on both sides of eqn.(\ref{eq:fthetascaling}) must now match. It is an elementary exercise to show that they indeed do. In summary, we have the result:
\be
\label{eq:fthetascalingfin}
L^{(s)}\,\left (\frac{\xi^\theta}{\sin\theta}\right )=0
\ee
{\bf (d) $\xi^r$}:

Eqn.(\ref{eq:boxxir}) implies
\be
\label{eq:rconjecture}
0 = \Box\,\xi^r\,-\,R^r_{\,\,r}\,\xi^r = (L^{(s)}_t\,+\,L^{(s)}_\theta\,+\,L^{(s)}_\phi)\xi^r+{\cal L}_r\,\xi^r
\ee
where
\bea
\label{eq:calLr}
{\cal L}_r &=& -\frac{1-kr^2}{R^2}\,\partial_r^2\,\xi^r\,+\,\left\{\frac{2(1-kr^2)}{R^2\,r}\,-\,\frac{kr}{R^2}\right\}\,\partial_r\,\xi^r\nonumber\\
& & -\,\left\{\frac{2(1-kr^2}{R^2\,r^2}\,+\,\frac{k}{R^2}\,+\,\frac{2(kr)^2}{R^2\,(1-kr^2}\right\}\,\xi^r
\eea
As before, we look for a scaling function $f_r(r)$ such that
\be
\label{eq:frscaling}
f_r(r)\,L^{(s)}_r\,\frac{\xi^r}{f_r(r)} = {\cal L}_r\,\xi^r\quad\quad
f_r(r)\,L^{(s)}_{t,\theta,\phi}\,\frac{\xi^r}{f_r(r)} =  L^{(s)}_{t,\theta,\phi}\,\xi^r
\ee
Explicitly,
\bea
\label{eq:rhscalLr}
f_r,L^{(s)}_r\,\frac{\xi^r}{f_r}=&-&\left\{\frac{2(1-kr^2}{R^2\,r}-\frac{kr}{R^2}\right\}\left (\partial_r\,\xi^r\,-\,\frac{f_r^\prime}{f_r}\,\xi^r\right )\nonumber\\
&-&\frac{1-kr^2}{R^2}\,\left\{\partial_r^2\,\xi^r-2\,\frac{f_r^\prime}{f_r}\,\partial_r\,\xi^r\,+\,\left (-\frac{f_r^{\prime\prime}}{f_r}\,+\,
2\,\frac{{f_r^\prime}^2}{f_r^2}\right )\xi^r\,\right\}
\eea
Equating the coefficients of $\partial_r\,\xi^r$ of eqns.(\ref{eq:calLr}, \ref{eq:rhscalLr}),
\be
\label{eq:frscaling2}
2\frac{1-kr^2}{R^2\,r}-\frac{kr}{R^2} = 2\frac{1-kr^2}{R^2}\,\frac{f_r^\prime}{f_r}\,-\,\{2\,\frac{1-kr^2}{R^2\,r}-\frac{kr}{R^2}\}
\ee
Simplifying,
\be
\label{eq:frscaling3}
2\,\frac{1-kr^2}{R^2}\,\frac{f_r^\prime}{f_r}\, =2\,\,\{2\,\frac{1-kr^2}{R^2\,r}-\frac{kr}{R^2}\}
\ee
On recognising
\be
\label{eq:frscaling4}
2\,\frac{1-kr^2}{r}\,-kr = (1-kr^2)\,\left(\ln{r^2\,\sqrt{1-kr^2}}\right )_{,r}
\ee
One finds that the solution of eqn.(\ref{eq:frscaling3}) is
\be
\label{eq:frscaling5}
f_r(r) = r^2\,\sqrt{1-kr^2}
\ee
and, finally,
\be
\label{eq:frscalingfin}
L^{(s)}\,\left(\frac{\xi^r}{r^2\,\sqrt{1-kr^2}}\right )=0
\ee
As in the case of $\xi^\theta$, a consistency check for this solution is that the coefficients of $\xi^r$ in eqns.(\ref{eq:calLr}, \ref{eq:rhscalLr}) match automatically. A little algebra shows that they indeed do match.

\subsection{Maximally symmetric submanifolds}
What we have shown so far is that suitably scaled Killing vector components of FLRW space are zero modes of the four-dimensional scalar 
Laplacian associated with the corresponding metric. The four-manifold can be maximally symmetric only when $R(t)$ satisfies
eqn.(\ref{eq:maximal}). Clearly, when the four-manifold itself is maximally symmetric, all the Killing vectors are non-zero (see next
section),
and they are zero modes of the \emph{four-dimensional} scalar Laplacian as shown.

When $R(t)$ does not satisfy this special condition, as is the case with generic FLRW-spaces, this full maximal symmetry is broken. Nevertheless,
the three dimensional sub-manifold parametrised by $t=constant, r,\theta,\phi$ is still maximally symmetric, and $\xi^t=0$ in this case. Restricting 
to this subspace means setting all t-derivatives to zero, or equivalently, $L^{(s)}_t=0$. Then the four-dimensional scalar Laplacian of
eqn.(\ref{eq:scalarlap}) reduces to the scalar Laplacian for this three-manifold and our conjecture for the Killing vector components $\xi^r,\xi^\theta,
\xi^\phi$ of this maximally symmetric 3-manifold follows. An additional consistency for this comes from the t-independence of the Killing 
vectors of eqn.(\ref{eq:3dkvupper}). As choosing a particular time-slice is equivalent to putting $R(t)=constant=R$, the same holds for the lower component Killing vectors too.

We can continue likewise to the two-dimensional submanifold with coordinates $\theta, \phi$ which is also maximally symmetric, by fixing
both r and t to be constant. This is equivalent to setting both $L^{(s)}_r$ and $L^{(s)}_t$ to zero. Then the four dimensional scalar 
Laplacian reduces to the two-dimensional scalar Laplacian on the sphere.

\section{Schr\"odinger construction for maximally symmetric FLRW spaces.}
In this section we explicitly construct the Killing vectors for the maximally symmetric FLRW spaces by applying the elegant method, due
originally to Schr\"odinger \cite{schrobook}, of
embedding the four-dimensional FLRW space in \emph{flat} five-dimensional Minkowski space. This method, for cartesian coordinates, can
also be found in Weinberg's book \cite{weinbook}, where he describes it without any reference to the originator! As already stated, $R(t)$ for such manifolds
can not be arbitrary, but must satisfy the eqn.(\ref{eq:maximal}).
For $k=0$, we have the de Sitter case $R(t) = e^{t}$. For $k=1$, one can have $R(t) = \cosh t$. We analyse both these cases here.
The $k=-1$ case can be obtained from the $k=1$ case through the formal substitution $t\,\rightarrow\,i\,t$. Therefore, in that case,
the scale factor becomes $R(t) = \sin\,t$.

The starting point is the metric for flat Minkowski space in five dimensions:
\be
\label{eq:flatM5}
ds_5^2 = dx_5^2\,-dx_4^2\,-dx_3^2\,-dx_2^2\,-dx_1^2
\ee
{\bf (i) k=0}:
Schr\"odinger considers the four-dimensional submanifold, defined by
\be
\label{eq:schr4sub0}
x_1^2\,+\,x_2^2\,+\,x_3^2\,+x_4^2\,-\,x_5^2 = 1
\ee
embedded in the flat $M_5$. The four-manifold coordinates (t,x,y,z) are introduced as:
\be
\label{eq:schr4coord0}
x_4+x_5 = e^t\quad\quad\,x_1 = e^t\,x\quad\,x_2 = e^t\,y\quad\,x_3 = e^t\,z
\ee
The polar coordinates are introduced through
\be
\label{eq:polar}
z = r\,\cos\theta\quad\,y = r\,\sin\theta\,\cos\phi\quad\,x = r\,\sin\theta\,\sin\phi
\ee
It is easy to show that the metric induced on the four-submanifold is:
\be
\label{eq:fourmetric0}
ds_4^2 = dt^2 - e^{2t}\,\left ( dr^2\,+\,r^2\,d\theta^2\,+\,r^2\,\sin^2\theta\,d\phi^2\right )
\ee
This is indeed of the FLRW type with $k=0$ and $R(t) = e^t$.

The flat five-dimensional space here is maximally symmetric and has 15 isometries. Of those, we need the ones that also keep the 
hypersurface eqn.(\ref{eq:schr4sub0}) unchanged. The five space-time translations of $M_5$ do not maintain this. On the other hand, the four
Lorentz-transformations and the six rotations of $M_5$ indeed maintain the hypersurface equation, and they in all induce 10 isometries on
the four-submanifold. That being the maximum number of isometries for a four-manifold, they must exhaust all the required isometries. We
first give the infinitesimal forms of these transformations from the five-dimensional perspective. 

\noindent
{\bf (i) Lorentz transformations on $M_5$}:

\bea
\label{eq:m5lorentz}
\delta\,x_5 &=& \delta\chi_1\,x_1\quad\delta\,x_5 = \delta\chi_2\,x_2\quad\delta\,x_5 = \delta\chi_3\,x_3\quad\delta\,x_5 = \delta\chi_4\,x_4\nonumber\\
\delta\,x_1 &=& \delta\chi_1\,x_5\quad\delta\,x_2 = \delta\chi_2\,x_5\quad\delta\,x_3 = \delta\chi_3\,x_5\quad\delta\,x_4 = \delta\chi_4\,x_5
\eea
{\bf (ii) Rotations on $M_5$}:

\bea
\label{eq:m5rot}
\delta\,x_1 &=& \delta\omega_1\,x_2\quad\delta\,x_1 = \delta\omega_2\,x_3\quad\delta\,x_1 = \delta\omega_3\,x_4\quad\delta\,x_2 = \delta\omega_4\,x_3\quad\delta\,x_2 = \delta\omega_5\,x_4\quad\delta\,x_3 = \delta\omega_6\,x_4\nonumber\\
\delta\,x_2 &=& -\delta\omega_1\,x_1\quad\delta\,x_3 = -\delta\omega_2\,x_1\quad\delta\,x_4 = -\delta\omega_3\,x_1\quad
\delta\,x_3 = -\delta\omega_4\,x_2\quad\delta\,x_4 = -\delta\omega_5\,x_2\quad\delta\,x_4 = -\delta\omega_6\,x_3\nonumber\\
\eea

Combining all the transformations, one can write
\bea
\label{eq:fouriso0}
\delta\,x_1 &=& \delta\omega_1\,x_2\,+\,\delta\omega_2\,x_3\,+\,\delta\omega_3\,x_4\,+\,\delta\chi_1\,x_5\nonumber\\
\delta\,x_2 &=& -\delta\omega_1\,x_2\,+\,\delta\omega_4\,x_3\,+\,\delta\omega_5\,x_4\,+\,\delta\chi_2\,x_5\nonumber\\
\delta\,x_3 &=& -\delta\omega_2\,x_1\,+\,\delta\omega_6\,x_4\,-\,\delta\omega_4\,x_2\,+\,\delta\chi_3\,x_5\nonumber\\
\delta\,x_4 &=& -\delta\omega_5\,x_2\,-\,\delta\omega_6\,x_3\,-\,\delta\omega_3\,x_1\,+\,\delta\chi_4\,x_5\nonumber\\
\delta\,x_5 &=& \delta\chi_1\,x_1\,+\,\delta\chi_2\,x_2\,+\,\delta\chi_3\,x_3\,+\,\delta\chi_4\,x_4
\eea

We now show how to calculate the Killing vectors in this method. Let us consider $\xi^t\equiv\,\delta t$.
\be
\label{eq:xitkeq0}
\delta(x_4+x_5)\,=\,e^t\,\delta t = (\delta\chi_1-\delta\omega_3)\,x_1\,+\,(\delta\chi_2-\delta\omega_5)\,x_2\,+\,(\delta\chi_3-\delta\omega_6)
\,x_3\,+\delta\chi_4\,(x_4+x_5)
\ee
On introducing
\bea
\label{eq:reparametrisek0}
\delta\chi_1\,-\,\delta\omega_3 &=& 2\,\delta a_x\quad\quad \delta\chi_1\,+\,\delta\omega_3 = 2\,\delta b_x\nonumber\\
\delta\chi_2\,-\,\delta\omega_5 &=& 2\,\delta a_y\quad\quad \delta\chi_2\,+\,\delta\omega_5 = 2\,\delta b_y\nonumber\\
\delta\chi_3\,-\,\delta\omega_6 &=& 2\,\delta a_z\quad\quad \delta\chi_3\,+\,\delta\omega_6 = 2\,\delta b_z\nonumber\\
\delta\chi_4 = \delta a_t
\eea
It follows that
\be
\label{eq:xitk0}
\xi^t = 2\,r\,(\delta a_x\,\sin\theta\,\sin\phi\,+\,\delta a_y\,\sin\theta\,\cos\phi\,+\,\delta a_z\,\cos\theta)\,+\,\delta a_t
\ee
Likewise, the Killing vector $\xi^r$ is obtained on first noting
\be
\label{eq:rsq}
r^2 = x^2\,+|,y^2\,+\,z^2\,=\,e^{-2t}\,(x_1^2\,+|,x_2^2\,+\,x_3^2)
\ee
Hence
\be
\label{eq:xirk01}
2\,r\,\delta\,r = -2\,e^{-2t}\,\delta\,t\,(x_1^2\,+\,x_2^2\,+\,x_3^2)\,+\,2\,e^{-2t}\,(x_1\,\delta\,x_1\,+\,x_2\,\delta\,x_2\,+\,\delta\,x_3\,x_3)
\ee
Using eqn.(\ref{eq:fouriso0}) it is seen that
\be
\label{eq:xirk02}
x_1\,\delta\,x_1\,+\,x_2\,\delta\,x_2\,+\,x_3\,\delta\,x_3 = (\delta\omega_3\,x_1\,+\,\delta\omega_5\,x_2\,+\,\delta\omega_6\,x_3)x_4\,+\,
(\delta\chi_1\,x_1\,+\,\delta\chi_2\,x_2\,+\,\delta\chi_3\,x_3)\,x_5
\ee
On using eqn.(\ref{eq:reparametrisek0}) and putting together one finds
\bea
\label{eq:xirk0}
\xi^r &=& -r\,\delta\,a_t\,+\,(\sin\theta\,\sin\phi\,\delta\,b_x\,+\,\sin\theta\,\cos\phi\,\delta\,b_y\,+\,\cos\theta\,\delta\,b_z)\,\nonumber\\
& & -\,(\sin\theta\,\sin\phi\,\delta\,a_x\,+\,\sin\theta\,\cos\phi\,\delta\,a_y\,+\,\cos\theta\,\delta\,a_z)\,(e^{-2t}\,+\,r^2)
\eea
The Killing vector $\xi^\theta$ is obtained from
\be
\label{eq:xithetak0}
z = r\,\cos\,\theta\,\rightarrow\,-r\,\sin\,\theta\,\delta\,\theta = \delta\,(e^{-t}\,x_3)\,-\,\delta\,r\,\cos\,\theta
\ee
We leave out the details by only giving the final result:
\bea
\label{eq:xithetak02}
\xi^\theta &=& -\,\frac{\sin\theta}{r}\,\delta\,b_z\,-\,(r-\frac{e^{-2t}}{r})\,\sin\theta\,\delta\,a_z\,+\,(r\,-\,\frac{e^{-2t}}{r})\,
(\cos\phi\,\delta\,a_y\,+\,\sin\phi\,\delta\,a_x)\,\cos\theta\nonumber\\
& &+\,\frac{\cos\theta}{r}\,(\cos\phi\,\delta\,b_y\,+\,\sin\phi\,\delta\,\delta\,b_x)\,+\,\sin\phi\,\delta\,\omega_2\,+\,\cos\phi\,\delta\,\omega_4
\eea
Finally, $\xi^\phi$ is computed thus:
\be
\label{eq:xiphik0}
\tan\phi\, = \frac{x}{y}=\frac{x_1}{x_2}\,\rightarrow\,\frac{\delta\,\phi}{\cos^2\,\phi}\,=\frac{x_2\,\delta\,x_1\,-\,x_1\,\delta\,x_2}{x_2^2}
\ee
Leaving out the details, the final result is:
\bea
\label{eq:xiphik02}
\xi^\phi &=& (\delta\omega_2\,\cos\phi\,-\,\delta\omega_4\,\sin\phi)\,\cot\theta\,\,+\,\frac{1}{r\,\sin\theta}\,
(\cos\phi\,\delta\,b_x\,-\,\sin\phi\,\delta\,b_y)\nonumber\\
& & +\frac{(e^{-2t}\,-\,r^2)}{r\,\sin\theta}\,(\sin\phi\,\delta\,a_y\,-\,\cos\phi\,\delta\,a_x)\,+\,\delta\omega_1
\eea
In summary, the Killing vectors for $k=0$ maximally symmetric FLRW space, as obtained by the Schr\"odinger embedding method are:
\begin{subequations}
\label{eq:k0schrkv}
\bea
\xi^t_{k=0} &=& 2\,r\,(\delta a_x\,\sin\theta\,\sin\phi\,+\,\delta a_y\,\sin\theta\,\cos\phi\,+\,\delta a_z\,\cos\theta)\,+\,\delta a_t\label{eq:k0schrkv-t}\\
\xi^r_{k=0} &=& -r\,\delta\,a_t\,+\,(\sin\theta\,\sin\phi\,\delta\,b_x\,+\,\sin\theta\,\cos\phi\,\delta\,b_y\,+\,\cos\theta\,\delta\,b_z)\nonumber\\
& & -\,(\sin\theta\,\sin\phi\,\delta\,a_x\,+\,\sin\theta\,\cos\phi\,\delta\,a_y\,+\,\cos\theta\,\delta\,a_z)\,(e^{-2t}\,+\,r^2)\label{eq:k0schrkv-r}\\
\xi^\theta_{k=0} &=& -\,\frac{\sin\theta}{r}\,\delta\,b_z\,-\,(r-\frac{e^{-2t}}{r})\,\sin\theta\,\delta\,a_z\,+\,(r\,-\,\frac{e^{-2t}}{r})\,
(\cos\phi\,\delta\,a_y\,+\,\sin\phi\,\delta\,a_x)\,\cos\theta\nonumber\\
& &+\,\frac{\cos\theta}{r}\,(\cos\phi\,\delta\,b_y\,+\,\sin\phi\,\delta\,\delta\,b_x)\,+\,\sin\phi\,\delta\,\omega_2\,+\,\cos\phi\,\delta\,\omega_4\label{eq:k0schrkv-theta}\\
\xi^\phi_{k=0} &=& (\delta\omega_2\,\cos\phi\,-\,\delta\omega_4\,\sin\phi)\,\cot\theta\,\,+\,\frac{1}{r\,\sin\theta}\,
(\cos\phi\,\delta\,b_x\,-\,\sin\phi\,\delta\,b_y)\nonumber\\
& & +\frac{(e^{-2t}\,-\,r^2)}{r\,\sin\theta}\,(\sin\phi\,\delta\,a_y\,-\,\cos\phi\,\delta\,a_x)\,+\,\delta\omega_1\label{eq:k0schrkv-phi}
\eea
\end{subequations}

{\bf (i) k=1}: Now we derive the analogous Killing vectors for the $k=1$ maximally symmetric case by using the Schr\"odinger method.
The equation for the embedded four-dimensional manifold is now taken to be:

\be
\label{eq:schr4sub1}
x_5^2\,-x_4^2-\,x_3^2\,-x_2^2\,-x_1^2 = 1
\ee
Note the important sign difference between eqn.(\ref{eq:schr4sub0}) and eqn.(\ref{eq:schr4sub1}). The coordinates $(t,r,\theta,\phi)$
on this hypersurface are introduced in a very different way from what was done for the $k=0$ case:
\bea
\label{eq:schr4coord1}
x_5\,&=&\,\sinh t\nonumber\\
x_4\,&=&\,\cosh t\,\cos\chi\nonumber\\
x_3\,&=&\,\cosh t\,\sin\chi\,\cos\theta\nonumber\\
x_2\,&=&\,\cosh t\,\sin\chi\sin\theta\cos\phi\nonumber\\
x_1\,&=&\,\cosh t\sin\chi\sin\theta\sin\phi
\eea
With $\sin\chi = r$, the induced metric on the hypersurface described by eqn.(\ref{eq:schr4sub1}) is:
\be
\label{eq:sub1metric}
ds_4^2 = dt^2\,-\,\cosh\,^2\,t\,(\frac{dr^2}{1-r^2}\,+\,r^2\,d\theta^2\,+\,r^2\,\sin^2\theta\,d\phi^2)
\ee
which is FLRW metric with $k=1$ and $R(t) = \cosh t$.

The isometries of this four dimensional space are found by the same method as in the $k=0$ case. We just give the final result 
for the relevant fivedimensional isometries( they are of course identical to eqn.(\ref{eq:fouriso0})):
\bea
\label{eq:fouriso1}
\delta\,x_1 &=& \delta\omega_1\,x_2\,+\,\delta\omega_2\,x_3\,+\,\delta\omega_3\,x_4\,+\,\delta\chi_1\,x_5\nonumber\\
\delta\,x_2 &=& -\delta\omega_1\,x_2\,+\,\delta\omega_4\,x_3\,+\,\delta\omega_5\,x_4\,+\,\delta\chi_2\,x_5\nonumber\\
\delta\,x_3 &=& -\delta\omega_2\,x_1\,+\,\delta\omega_6\,x_4\,-\,\delta\omega_4\,x_2\,+\,\delta\chi_3\,x_5\nonumber\\
\delta\,x_4 &=& -\delta\omega_5\,x_2\,-\,\delta\omega_6\,x_3\,-\,\delta\omega_3\,x_1\,+\,\delta\chi_4\,x_5\nonumber\\
\delta\,x_5 &=& \delta\chi_1\,x_1\,+\,\delta\chi_2\,x_2\,+\,\delta\chi_3\,x_3\,+\,\delta\chi_4\,x_4
\eea
The calculation of $\xi^t$ in this case is different than in the $k=0$ case:
\be
\label{eq:xitk1}
\delta\,x_5 = \cosh\,t\,\delta\,t = \cosh\,t\,\{\delta\chi_1\,r\,\sin\theta\sin\phi\,+\,\delta\chi_2\,r\,\sin\theta\cos\phi\,+\,\delta\chi_3\,r\,\cos\theta \,+\,\delta\chi_4\,\sqrt{1-r^2}\}
\ee
Therefore
\be
\label{eq:xitk12}
\xi^t_{k=1} = \delta\chi_1\,r\,\sin\theta\sin\phi\,+\,\delta\chi_2\,r\,\sin\theta\cos\phi\,+\,\delta\chi_3\,r\,\cos\theta \,+\,\delta\chi_4\,\sqrt{1-r^2}
\ee
All other Killing vectors are calculated exactly as before. We summarise the results without giving the details.

\begin{subequations}
\label{eq:k1schrkv}
\bea
\xi^t_{k=1} &=& \delta\chi_1\,r\,\sin\theta\sin\phi\,+\,\delta\chi_2\,r\,\sin\theta\cos\phi\,+\,\delta\chi_3\,r\,\cos\theta \,+\,\delta\chi_4\,\sqrt{1-r^2}\label{eq:k1schrkv-t}\\
\xi^r_{k=1}\,&=&\,\frac{{\dot R(t)}}{R(t)}(1-r^2)\,\big\{\delta\chi_1\,\sin\theta\sin\phi+\delta\chi_2\,\sin\theta\cos\phi+\delta\chi_3\,\cos\theta\big\}-\frac{{\dot R(t)}}{R(t)}\,\delta\chi_4\,r\sqrt{1-r^2}\nonumber\\
& &\,+\sqrt{1-r^2}\,\big\{\delta\omega_6\,\cos\theta+\delta\omega_3\,\sin\theta\sin\phi+\delta\omega_5\sin\theta\cos\phi\big\}\label{eq:k1schrkv-r}\\
\xi^\theta_{k=1}\,&=&\,-\frac{\sqrt{1-r^2}}{r}\,\big\{\delta\omega_6\,\sin\theta-\delta\omega_5\,\cos\theta\cos\phi-\delta\omega_3\,\cos\theta\sin\phi\big\}\nonumber\\
& &\,-\big\{-\delta\omega_4\,\cos\phi-\delta\omega_2\,\sin\phi\big\}\nonumber\\
& &\,-\frac{{\dot R(t)}}{R(t)}\,\frac{1}{r}\,\big\{\delta\chi_3\,\sin\theta-\delta\chi_1\,\cos\theta\sin\phi-\delta\chi_2\,\cos\theta\cos\phi\big\}\label{eq:k1schrkv-theta}\\
\xi^\phi_{k=1}\,&=&\,\frac{{\dot R(t)}}{R(t}\,\frac{1}{r}\,\big\{\delta\chi_2\,\frac{\sin\phi}{\sin\theta}-\delta\chi_1\,\frac{\cos\phi}{\sin\theta}\big\}\nonumber\\
& &\,-\frac{\sqrt{1-r^2}}{r}\,\big\{\delta\omega_3\,\frac{\cos\phi}{\sin\theta}-\delta\omega_5\,\frac{\sin\phi}{\sin\theta}\big\}\nonumber\\
& &\,-\big\{\delta\omega_1\,-\delta\omega_4\,\frac{\cos\theta}{\sin\theta}\,\sin\phi+\delta\omega_2\,\frac{\cos\theta}{\sin\theta}\,\cos\phi\big\}\label{eq:k1schrkv-phi}
\eea
\end{subequations}
\section{Some consequences}
In this section we work out some consequences of our result. Let us consider eqn.(\ref{eq:3dkvupper-phi}) first; the $\theta,\phi$ dependences
of the various independent $\xi^\phi$ satisfy
\be
\label{eq:3dphiconsequence}
\left ({\cal D}_\phi^2\,+\,{\cal D}_\theta^2\right )\,f(\theta,\phi) =0\,\quad\rightarrow L^{(s)}_{\theta,\phi}\,\xi^\phi = 0
\ee
But as per our results, the entire $\xi^\phi$ must satisfy
\be
\label{eq:3dphiconsequence2}
\left( L^{(s)}_r\,+\,L^{(s)}_{\theta,\phi}\,\right ) \xi^\phi = 0
\ee
Consequently, the radial parts must satisfy
\be
\label{eq:3dphiconsequence3}
L^{(s)}_r\,f_\phi(r) = 0
\ee
The two radial dependences in eqn.(\ref{eq:3dkvupper-phi}) are i) $f_\phi(r) = constant$, and, $f_\phi(r) = \frac{\sqrt{1-kr^2}}{r}$. On using
eqn.(\ref{eq:scalarlap-r}), it is seen that both of them indeed satisfy eqn.(\ref{eq:3dphiconsequence3}).

Next, let us turn to eqn.(\ref{eq:3dkvupper-r}). The angular dependences occurring in it are 
$f(\theta,\phi) = \sin\theta\,\cos\phi, \sin\theta\,\sin\phi, \cos\theta$. None of them satisfies eqn.(\ref{eq:3dphiconsequence}). 
However, all of them satisfy
\be
\label{eq:3drconsequence}
L^{(s)}_{\theta,\phi}\,f(\theta,\phi) = \frac{2}{R^2\,r^2}\,\,f(\theta,\phi)
\ee
But as per our results
\be
\label{eq:3drconsequence2}
\left (L^{(s)}_r\,+\,L^{(s)}_{\theta,\phi}\right )\left (\frac{\xi^r}{r^2\,\sqrt{1-kr^2}}\right ) =0
\ee
which translates, in the present context to
\be
\label{eq:3drconsequence}
L^{(s)}_r\,\frac{1}{r^2} = -\,\frac{2}{R^2}\,\frac{1}{r^4}
\ee
where use has been made of eqn.(\ref{eq:3drconsequence}). This is indeed seen to be satisfied.

Finally, let us consider the Killing vector of eqn.(\ref{eq:k1schrkv-t}). $k=1$ now. There are two types involved. Let us first look at the type
with non-trivial $\theta, \phi$ dependence. The angular dependence is exactly of the type considered in the previous case, and hence
eqn.(\ref{eq:3drconsequence}) is satisfied here too. Just as in that case, there is no t-dependence here too. On the other hand, the radial
dependence now is $f_t(r)=r$, while in the previous case it was $\sqrt{1-kr^2}$. But our results require that it is this Killing vector
scaled by $\frac{1}{R^3(t)}$ that must be a zero mode of the $L^{(s)}$. This requires
\be
\label{eq:4dtconsequence}
L^{(s)}_{t,r}\,\frac{r}{R^3(t)} = -\,\frac{2}{R^2(t)\,r^2}\,\frac{r}{R^3(t)}
\ee
This can be checked to be true on noting
\be
\label{eq:4dtconsequence2}
L^{(s)}_r\,r = -\frac{2-3r^2}{R^2(t)\,r}
\ee
and,
\be
\label{eq:4dtconsequence3}
L^{(s)}_t\,\frac{1}{R^3(t)} = -\frac{3}{R^5(t)}
\ee
The second type of Killing vector in eqn.(\ref{eq:k1schrkv-t}) has no $(t,\theta,\phi)$ dependences but has only a $\sqrt{1-r^2}$
radial dependence. Hence in this case
\be
\label{eq:4dtconsequence2}
L^{(s)}_{\theta,\phi}\,\xi^t_{k=1} = 0
\ee
This requires, considering the $R^{-3}$-scaling needed in this case also,
\be
\label{eq:4dtconsequence4}
L^{(s)}_{t,r}\,\frac{\sqrt{1-r^2}}{R^3(t)} =0
\ee
Indeed, it is straightforward to verify this on using eqn.(\ref{eq:4dtconsequence3}), and,
\be
\label{eq:4dtconsequence5}
L^{(s)}_r\,\sqrt{1-r^2} = \frac{3}{R^2}\,\sqrt{1-r^2}
\ee
\section{Discussion}
We have uncovered a new property of Killing vectors of FLRW-metric in comoving coordinates, namely, their components being 
zero modes of the scalar Laplacian upon suitable scaling. We saw in the last section that this adds another perspective to 
the actual solutions of the Killing equations. Reversing
the logic of the previous section, it may be possible to provide a transparent method of finding the explicit solutions. This will be taken up
elsewhere. We expect similar results to hold for higher-dimensional Laplacians, at least to the higher-dimensional variants of the FLRW-metric.
Higher-dimensional Laplacians play a central role in many areas like Supergravity theories, physics of extra dimensions, braneworld scenarios, 
String theory etc.
In the context of AdS/CFT correspondence also, their properties play a key role. For example, in \cite{eastwood}, the mathematical problem
of the symmetries of Laplacians is investigated from this perspective. Zero modes also make explicit appearance in many 
diverse contexts. We cite here a few such.

The temperature-independent parts of the partition functions associated with $AdS_2$ have been shown to be related to the zero
modes \cite{keelerng}. Again, in the context of AdS/CFT conjecture, zero modes have been found and related to instabilities
of topological blackholes \cite{belinmaloney}. In the context of the so called braneworld scenarios based on the Randall-Sundrum model \cite{rsundrum} with noncompact fifth dimension, the so called Gravi-scalar modes were found to have zero modes \cite{durrerkocian}.
These authors even proposed experimental tests in binary pulsar systems. 
In a somewhat different context, Eastwood \cite{eastwood}, motivated by the symmetry of higher-spin field theories, has made a general study
of symmetries of Laplacians. It will be interesting to study the mutual interplay between such symmetries and our results. 

Killing vectors arise in such diverse contexts, as for example in \cite{boyer}, and in supergravity and superstring theories. An immediate
generalization of the concept of Killing vectors, namely, Killing Tensors, are also important in a variety of contexts. For example, they
play an important role for Kerr-geometries \cite{carter}. They are also important in determining constants of motion \cite{sommers}. The
equations for Killing tensors are much harder to solve. In $d=4$, they involve 20 PDE's as compared to the 10 for Killing vectors. Another
important generalization is that of Conformal-Killing vectors, and likewise Conformal-Killing tensors. Finally, in Supersymmetric theories
like Supergravity and Superstring theories, the so called Killing-spinors play important roles \cite{residual}. The integrability conditions
for such spinors require properties of Laplacians.
Therefore there are a number of interesting contexts where our results
may have potential uses. 


\end{document}